\documentclass[prd,notitlepage,superscriptaddress,nofootinbib,preprintnumbers]{revtex4-2}

\usepackage{amsmath}
\usepackage{amsfonts}
\usepackage{extarrows}
\usepackage{graphicx}
\usepackage[utf8]{inputenc}
\usepackage{slashed}
\usepackage{xcolor}

\usepackage{hyperref}
\hypersetup{colorlinks, linkcolor = [rgb]{0,0.0,0.75}, citecolor = [rgb]{0,0.0,0.75}, urlcolor = [rgb]{0,0.0,0.75}}

\makeatletter
\g@addto@macro\bfseries{\boldmath}
\makeatother

\DeclareMathOperator{\Tr}{Tr}
\DeclareMathOperator{\sgn}{sgn}
\newcommand{\MSbar}{\overline{\text{MS}}}
\newcommand{\MMS}{\text{M}\MSbar}
\newcommand{\RI}{\text{RI}}
\newcommand{\cO}{\mathcal{O}}
\newcommand{\vvx}{\boldsymbol{x}}
\newcommand{\vvp}{\boldsymbol{p}}
\newcommand{\vvzhat}{\boldsymbol{\hat z}}
\newcommand{\vvzero}{\boldsymbol{0}}
\newcommand{\zmax}{z_\text{max}}

\begin{document}

\preprint{CERN-TH-2020-181}

\title{Lattice continuum-limit study of nucleon quasi-PDFs}

\author{Constantia~Alexandrou}
\affiliation{Department of Physics, University of Cyprus, P.O.\ Box 20537, 1678 Nicosia, Cyprus}
\affiliation{Computation-based Science and Technology Research Center, The Cyprus Institute, 20 Kavafi Str., Nicosia 2121, Cyprus}
\author{Krzysztof~Cichy}
\affiliation{Faculty of Physics, Adam Mickiewicz University, Uniwersytetu Poznańskiego 2, 61-614 Poznań, Poland}
\author{Martha~Constantinou}
\affiliation{Temple University, 1925 N.\ 12th Street, Philadelphia, PA 19122-1801, USA}
\author{Jeremy~R.~Green}
\affiliation{Theoretical Physics Department, CERN, 1211 Geneva 23, Switzerland}
\author{Kyriakos~Hadjiyiannakou}
\affiliation{Department of Physics, University of Cyprus, P.O.\ Box 20537, 1678 Nicosia, Cyprus}
\affiliation{Computation-based Science and Technology Research Center, The Cyprus Institute, 20 Kavafi Str., Nicosia 2121, Cyprus}
\author{Karl~Jansen}
\affiliation{NIC, Deutsches Elektronen-Synchrotron, 15738 Zeuthen, Germany}
\author{Floriano~Manigrasso}
\affiliation{Department of Physics, University of Cyprus, P.O.\ Box 20537, 1678 Nicosia, Cyprus}
\affiliation{Institut für Physik, Humboldt-Universität zu Berlin, Newtonstr.\ 15, 12489 Berlin, Germany}
\affiliation{Dipartimento di Fisica, Università di Roma ``Tor Vergata'', Via della Ricerca Scientifica 1, 00133 Rome, Italy}
\author{Aurora~Scapellato}
\affiliation{Faculty of Physics, Adam Mickiewicz University, Uniwersytetu Poznańskiego 2, 61-614 Poznań, Poland}
\author{Fernanda~Steffens}
\affiliation{Institut für Strahlen- und Kernphysik, Universität Bonn, Nussallee 14--16, 53115 Bonn, Germany}

\date{\today}

\begin{abstract}
  The quasi-PDF approach provides a path to computing parton
  distribution functions (PDFs) using lattice QCD. This
  approach requires matrix elements of a power-divergent operator in a
  nucleon at high momentum and one generically expects
  discretization effects starting at first order in the lattice
  spacing $a$. Therefore, it is important to demonstrate that the continuum limit
  can be reliably taken and to understand the size and shape of
  lattice artifacts. In this work, we report a calculation of
  isovector unpolarized and helicity PDFs using lattice ensembles with
  $N_f=2+1+1$ Wilson twisted mass fermions, a pion mass of
  approximately 370~MeV, and three different lattice spacings. Our
  results show a significant dependence on $a$, and the continuum
  extrapolation produces a better agreement with phenomenology. The
  latter is particularly true for the antiquark distribution at small
  momentum fraction $x$, where the extrapolation changes its sign.
\end{abstract}

\maketitle

\section{Introduction}

The calculation of parton distribution functions (PDFs) using lattice
QCD has seen renewed interest in recent years~\cite{Monahan:2018euv,
  Cichy:2018mum,Ji:2020ect,Constantinou:2020pek}, driven in part by the introduction of the quasi-PDF
method~\cite{Ji:2013dva,Ji:2014gla}. This method requires nucleon matrix elements
of a nonlocal operator containing a Wilson line, which must be
computed on the lattice. Previous calculations of quasi-PDFs and
related observables using the same operator by ETMC are given in
Refs.~\cite{Alexandrou:2015rja,Alexandrou:2016jqi,
  Alexandrou:2017huk,Alexandrou:2018pbm,Alexandrou:2018eet,
  Alexandrou:2019lfo,Alexandrou:2019dax,Chai:2020nxw,
  Alexandrou:2020zbe} and by other collaborations in
Refs.~\cite{Lin:2014zya,Chen:2016utp,Zhang:2017bzy,Chen:2017mzz,
  Orginos:2017kos,Lin:2017ani,Chen:2017lnm,Chen:2017gck,
  Chen:2018xof,Chen:2018fwa,Lin:2018qky,Karpie:2018zaz,
  Liu:2018hxv,Chen:2019lcm,Izubuchi:2019lyk,Cichy:2019ebf,Joo:2019jct,
  Joo:2019bzr,Lin:2020ssv,Joo:2020spy,Bhattacharya:2020cen,Bhattacharya:2020xlt,Bhattacharya:2020jfj,Zhang:2020dkn,Bhat:2020ktg,Fan:2020nzz,Gao:2020ito,Bringewatt:2020ixn,DelDebbio:2020rgv}.

The presence of a Wilson line in the nonlocal operator introduces a
power divergence. This divergence must be exactly removed by the
renormalization procedure so that a finite continuum limit can be
obtained. Furthermore, in contrast to the case of local operators, the
use of a lattice action with exact chiral symmetry or at maximal twist
does not eliminate all discretization effects linear in the lattice
spacing $a$~\cite{Green:2017xeu, Chen:2017mie, Green:2020xco}. This
means that in a lattice setup where most observables have only
$O(a^2)$ lattice artifacts, quasi-PDFs can nevertheless have $O(a)$
contributions. For both of these reasons, it is important to
numerically study the approach to the continuum limit so that future
calculations will be better equipped to control all sources of
systematic uncertainty.

There exist some previous studies using more than one lattice
spacing. Ref.~\cite{Green:2017xeu} includes an early analysis using
two of the three lattice spacings used in this work. Nonperturbative
renormalization was studied using two lattice spacings in
Ref.~\cite{Izubuchi:2019lyk}, and the same two lattice spacings were
used for studying pion PDFs in Ref.~\cite{Gao:2020ito}. After the
first version of this paper was submitted, two more works appeared.
Ref.~\cite{Lin:2020fsj} presents a study of nucleon PDFs using three
lattice spacings and three different pion masses, in which the lowest
two pion masses were each studied using a single lattice spacing and
the highest pion mass was studied using two lattice spacings. Finally,
zero-momentum pion matrix elements were computed in
Ref.~\cite{Zhang:2020rsx} using multiple actions and up to four
lattice spacings per action.

In this paper, we present a study of the approach to the continuum
limit of isovector nucleon unpolarized and helicity parton
distributions using three lattice ensembles, each having a different
lattice spacing but with otherwise similar parameters.
Section~\ref{sec:setup} describes the ensembles and
the observables we compute. A dedicated study on one ensemble of
systematic effects from excited-state contamination is reported in
Section~\ref{sec:excited_states}. Renormalization factors are obtained
using two different methods in Section~\ref{sec:renormalization}; in
addition, we study a ratio of matrix elements that cancels the
renormalization. In Section~\ref{sec:continuum}, we take the continuum
limit, both for position-space matrix elements and for PDFs. Finally,
conclusions are given in Section~\ref{sec:conclusions}.

\section{Lattice setup}
\label{sec:setup}

\begin{table}
  \centering
  \begin{tabular}{l|ccc|cc|cccc|cc}
    Name & $\beta$ & $a\mu_l$ & size & $a$ (fm) & $m_\pi$ (MeV) & $p_z L/(2\pi)$ & $p_z$ (GeV) & $t_s/a$ & $t_s$ (fm) & $N_\text{conf}$ & $N_\text{samp}$\\\hline
    A60 & 1.90 & 0.0060 & $24^3\times 48$ & 0.0934(13)(35) & 365 & 3 & 1.66 & 10 & 0.934 & 1260 & 40320 \\
    B55 & 1.95 & 0.0055 & $32^3\times 64$ & 0.0820(10)(36) & 373 & 4 & 1.89 & 12 & 0.984 & 1829 & 58528 \\
    D45 & 2.10 & 0.0045 & $32^3\times 64$ & 0.0644(07)(25) & 371 & 3 & 1.80 & 15 & 0.966 & 1259 & 40288 \\
  \end{tabular}
  \caption{Parameters of the three $N_f=2+1+1$ lattice ensembles: gauge
    coupling $\beta$, bare light quark mass $a\mu_l$, and size. The pion
    mass $m_\pi$ and lattice spacing $a$ (determined via the nucleon mass)
    are taken from Ref.~\cite{Alexandrou:2013joa}. Nucleon three-point
    functions are computed with momentum $\vec p=(0,0,p_z)$ and source-sink
    time separation $t_s$. The total number of gauge configurations is given by
    $N_\text{conf}$; on each one, we use an evenly-spaced grid of 32 source
    positions, with a random overall displacement, yielding $N_\text{samp} =
    32 N_\text{conf}$ samples.}
  \label{tab:ensembles}
\end{table}

We use three lattice ensembles that differ primarily in their lattice
spacings $a=0.0644$, 0.0820, and 0.0934~fm. These have dynamical
degenerate up and down quarks with pion mass approximately 370~MeV and
dynamical strange and charm quarks with near-physical masses, i.e.\
$N_f=2+1+1$. The gauge action is Iwasaki~\cite{Iwasaki:1985we,
  Iwasaki:2011np} and the fermions use Wilson twisted mass tuned to
maximal twist. These ensembles were generated by
ETMC~\cite{Baron:2010bv}; parameters for the three used in this work
are given in Table~\ref{tab:ensembles}. The ensemble with intermediate
lattice spacing, B55, was previously used by some of us for studying
quasi-PDFs in Refs.~\cite{Alexandrou:2015rja, Alexandrou:2016jqi,
  Alexandrou:2017huk}.

Isovector quasi-PDFs are obtained from nucleon matrix elements of the
nonlocal operator
\begin{equation}
  \cO_\Gamma(\vvx,z) = \bar\psi(\vvx+z\vvzhat) \Gamma \tau_3 W(\vvx+z\vvzhat,\vvx) \psi(\vvx),
  \label{eq:def_operator}
\end{equation}
where bold symbols denote Euclidean four-vectors, $\psi$ is the
doublet of light quarks, $W$ is a Wilson line, $\tau_3$ selects the
isovector $u-d$ flavor combination, and we have chosen to extend the
operator in the third spatial direction. We employ five steps of stout
smearing~\cite{Morningstar:2003gk} in the definition of $W$. The
operator's nucleon matrix elements can be written as
\begin{equation}
  \left\langle \vvp,s' \middle| \cO_\Gamma(\vvzero,z;\mu) \middle| \vvp,s \right\rangle
  = h_\Gamma(p_z,z;\mu) \bar u(\vvp,s') \Gamma u(\vvp,s),
\end{equation}
where $\mu$ represents the scale at which $\cO$ is
renormalized. Taking the Fourier transform, we obtain the unpolarized
and helicity quasi-PDFs,
\begin{equation}
\begin{aligned}\label{eq:quasi-PDF}
\tilde q(x,p_z;\mu) &=\frac{p_z}{2\pi} \int dz\, e^{-i x p_z z} h_{\gamma_0}(p_z,z;\mu),\\
\Delta\tilde q(x,p_z;\mu) &= \frac{p_z}{2\pi} \int dz\, e^{-i x p_z z} h_{\gamma_3\gamma_5}(p_z,z;\mu).
\end{aligned}
\end{equation}
These are related to physical PDFs through factorization,
\begin{equation}
  \tilde{q}(x,p_z;\mu) = \int \frac{d\xi}{|\xi|} C\left(\xi,\frac{\mu}{p_z}\right)
  q\left(\frac{x}{\xi};\mu\right)
  + O\left(\frac{m_N^2}{p_z^2}, \frac{\Lambda_\text{QCD}^2}{p_z^2}\right),
\end{equation}
and a similar expression applies to the helicity case.

The details of our calculation are similar to
Ref.~\cite{Alexandrou:2019lfo}, although we use nucleon momenta only
in the $+\hat z$ direction and do not improve statistics by averaging
over equivalent directions. The proton interpolating operator is
defined using Wuppertal-momentum-smeared quark
fields~\cite{Gusken:1989qx, Bali:2016lva}, with the smearing performed
using APE-smeared gauge links~\cite{Albanese:1987ds}.

\section{Excited-state effects}
\label{sec:excited_states}

\begin{table}
  \centering
  \begin{tabular}{c|rcr}
    $t_s/a$ & $N_\text{conf}$ & $N_\text{src}$ & $N_\text{samp}$ \\\hline
    $\{4,5,6,7\}$ & 315 & 4 & 1260 \\
    8 & 315 & 8 & 2520 \\
    9 & 315 & 16 & 5040 \\
    10&1260 & 32 & 40320 \\
  \end{tabular}
  \caption{Statistics used for excited-state study on ensemble A60.}
  \label{tab:A60_excited_states}
\end{table}

\begin{figure}
  \centering
  \includegraphics{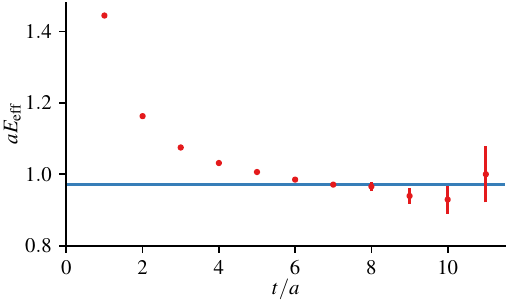}
  \caption{Nucleon effective energy on ensemble A60 with momentum
    $p_z=3(2\pi/L)$. The horizontal
    line indicates the predicted energy using the nucleon mass from
    Ref.~\cite{Alexandrou:2013joa} and the continuum dispersion
    relation.}
  \label{fig:A60_Eeff}
\end{figure}

On ensemble A60, we performed a dedicated study of excited-state
effects by varying the source-sink separation $t_s/a$ from 4 to 10. The
nucleon effective energy on this ensemble is shown in
Fig.~\ref{fig:A60_Eeff}; although momentum smearing yields a good
signal at moderate source-sink separations, the statistical
uncertainty still grows rapidly at large separations. Therefore, we
use much larger statistics for the larger separations, as given in
Table~\ref{tab:A60_excited_states}.

Matrix elements are obtained from two-point and three-point
correlation functions $C_\text{2pt}(t_s)$ and
$C_{\text{3pt}}^{\Gamma,z}(\tau,t_s)$, where $t_s$ is the Euclidean time
separation between the source and the sink and $\tau$ is the
Euclidean time separation between the source and $\cO_\Gamma(z)$. We consider two
estimators for the matrix element $h_\Gamma(z)$:
\begin{gather}
  \begin{aligned}
  h_{\Gamma,\text{eff}}^\text{ratio}(z;t_s)
  &\equiv \frac{C_\text{3pt}^{\Gamma,z}(\tfrac{t_s}{2},t_s)}{C_\text{2pt}(t_s)} \\
  &= h_\Gamma(z) + O\left(e^{-\Delta E t_s/2}\right), \\
  h_{\Gamma,\text{eff}}^\text{summ}(z;t_s)
  &\equiv \frac{S_{\Gamma,z}(t_s+a)-S_{\Gamma,z}(t_s)}{a} \\
  &= h_\Gamma(z) + O\left(e^{-\Delta E t_s}\right),
  \end{aligned} \\
 \text{where}\quad S_{\Gamma,z}(t_s) \equiv a\sum_{\tau/a=1}^{t_s/a-1} \frac{C_\text{3pt}^{\Gamma,z}(\tau,t_s)}{C_\text{2pt}(t_s)}
\end{gather}
and $\Delta E$ is the energy gap to the lowest excited state.

\begin{figure*}
  \centering
  \includegraphics[width=0.76\textwidth]{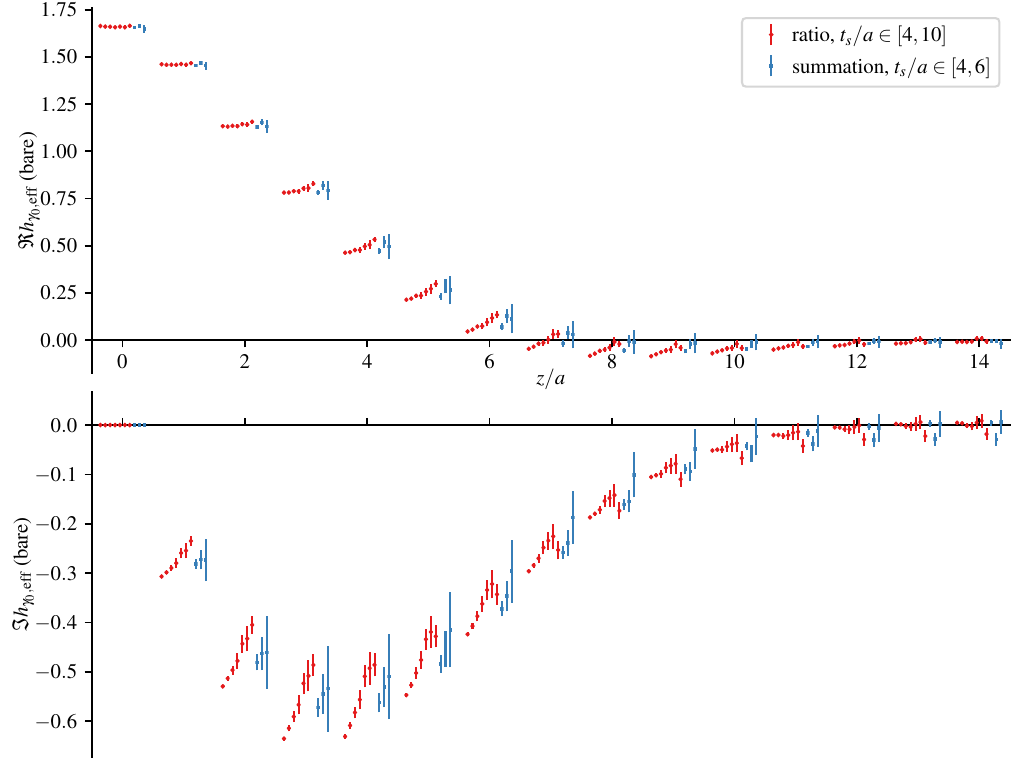}
  \caption{Effective $h_{\gamma_0}$ versus $z$: real part (top) and
    imaginary part (bottom). For each $z$, the ratio-midpoint results
    are shown using seven source-sink separations, increasing from
    left to right (red), and the summation-method results are shown
    using the shortest three source-sink separations (blue).}
  \label{fig:A60_excited_states_unpol}
\end{figure*}

\begin{figure*}
  \centering
  \includegraphics[width=0.76\textwidth]{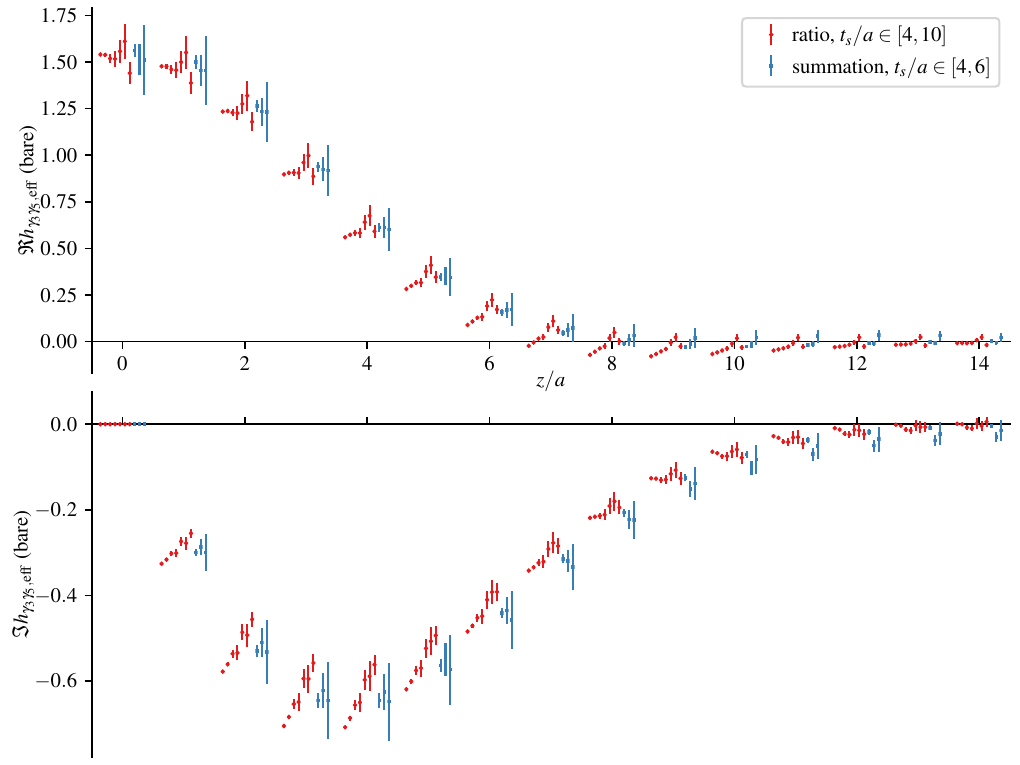}
  \caption{Effective $h_{\gamma_3\gamma_5}$ versus $z$. See the
    caption of Fig.~\ref{fig:A60_excited_states_unpol}.}
  \label{fig:A60_excited_states_hel}
\end{figure*}

Results are shown for the unpolarized and helicity matrix elements in
Figs.~\ref{fig:A60_excited_states_unpol} and
\ref{fig:A60_excited_states_hel}. For both observables the
excited-state effects are similar. In the real part at small $z$, the
dependence on $t_s$ is weak, especially for the unpolarized case where
$h_{\gamma_0}(0)$ is a conserved charge. For $z>6a$, $h_\text{eff}$
dips below zero at small $t_s$ and this negative part is substantially
reduced when $t_s$ is increased. In the imaginary part, the negative
peak around $z=3a$ is reduced in magnitude when $t_s$ is increased.

\begin{figure*}
  \centering
  \includegraphics[width=0.495\textwidth]{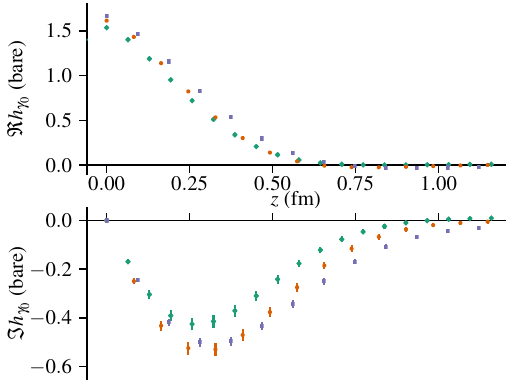}
  \includegraphics[width=0.495\textwidth]{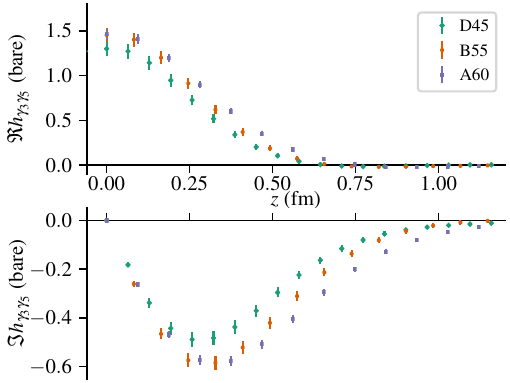}
  \caption{Bare matrix elements: real part (top) and imaginary part
    (bottom) for the unpolarized (left) and helicity (right)
    operators.}
  \label{fig:ME_bare}
\end{figure*}

For most values of $z$, $h_\text{eff}^\text{ratio}(z)$ with $t_s=10a$ is
consistent with the value for $t_s=8a$ and $9a$ and also with
$h_\text{eff}^\text{summ}(z)$ for $t_s=5a$ and $6a$. Therefore we
conclude that excited-state effects are reasonably under control using
the ratio method with the largest time separation, and we choose to use
similar separations for the two other ensembles. However, the analysis
in the rest of this paper differs slightly from the excited-state
study: instead of simply taking the midpoint $\tau=t_s/2$ in
$C_\text{3pt}^{\Gamma,z}(\tau,t_s)$, we average over several central
values of $\tau$ to reduce the statistical uncertainty. The resulting
bare matrix elements for all three ensembles are shown in
Fig.~\ref{fig:ME_bare}.

One risk of studying excited-state effects using just one ensemble is
that insufficiently controlled excited-state contributions on the
other ensembles could be mistakenly interpreted as discretization
effects. To reduce this possibility, $t_s$ was chosen to be slightly
larger on the two ensembles that lack an excited-states
study. Furthermore, our findings in Section~\ref{sec:p0ratio}, that
accounting for the leading effect of small differences in $p_z$
improves the approach to the continuum, and in
Section~\ref{sec:continuum}, that the dependence on $a$ is typically
monotonic, are both consistent with discretization effects and not
excited-state effects playing the dominant role in this study.

\section{Renormalization}
\label{sec:renormalization}

Renormalization of the nonlocal operator $\cO(z)$ was a stumbling
block in rigorously calculating quasi-PDFs and was absent in the
earliest lattice QCD calculations~\cite{Lin:2014zya,
  Alexandrou:2015rja, Chen:2016utp, Alexandrou:2016jqi}. In contrast
with local quark bilinears that diverge logarithmically, $\cO(z)$
contains a Wilson line that introduces a power divergence. In order to
obtain a continuum limit, it is essential that this divergence be
removed exactly, meaning that lattice perturbation theory is
inadequate. Nonperturbative renormalization
prescriptions~\cite{Alexandrou:2017huk, Chen:2017mzz, Green:2017xeu},
introduced more than three years after the first lattice quasi-PDF
calculations, are necessary.

We employ two different methods for nonperturbative renormalization,
both of which involve imposing renormalization conditions on Green's
functions evaluated on Landau-gauge-fixed lattices\footnote{A hybrid 
  approach that incorporates elements of both methods was recently
  proposed in Ref.~\cite{Ji:2020brr}.}. For this, we use
the $N_f=4$ twisted mass ensembles from Ref.~\cite{Alexandrou:2015sea}
listed in Table~\ref{Table:Z_ensembles}. These have the same action
and bare coupling as the ensembles used for computing nucleon matrix
elements. However, because of the difficulty in reaching maximal twist
with four degenerate light fermions, we instead average over pairs of
ensembles with opposite PCAC masses. After renormalizing
$\cO_\Gamma(z)$ in a nonperturbative intermediate scheme, perturbation
theory is used to convert first to the $\MSbar$ scheme and then to a
modified $\MSbar$ ($\MMS$) scheme~\cite{Alexandrou:2019lfo}. The
latter cancels a $\log(z^2)$ divergence in the $\MSbar$-renormalized
matrix element at short distance and enables a matching between
quasi-PDF and PDF that conserves charge.

The first method is the whole operator approach, where renormalization
conditions are imposed independently on $\cO_\Gamma(z)$ for each $z$,
producing a separate renormalization factor for each $z$. The
procedure is very similar to methods commonly used for local quark
bilinears, and the nonperturbative intermediate scheme is RI$'$-MOM.

The second method is the auxiliary field approach, where the nonlocal
operator is rewritten as a pair of local operators in an extended
theory. Renormalization conditions are imposed on those local
operators and on the action of the extended theory, producing a
minimal set of renormalization parameters. The nonperturbative
intermediate scheme, RI-xMOM, uses a mixture of momentum space and
position space.

When $z=0$, it is a special case where $\cO_\Gamma(z)$ is a local
operator, namely a vector or axial current. For this point, we use the
renormalization factor for the corresponding local operator determined
in Ref.~\cite{Alexandrou:2015sea}.

In the next two subsections we discuss each method and their sources
of systematic uncertainty. In a third subsection, we form a ratio of
nucleon matrix elements to cancel the renormalization of
$\cO_\Gamma(z)$ and study the continuum limit of the ratio.

\begin{table}
\begin{center}
\begin{tabular}{cccccc}
\hline
\hline
$a \mu$ & $\kappa$ & $a \mu^\text{sea}_{\rm PCAC}$ & $a M_{PS}$ & lattice size & used in\\
\hline
\hline
$\,\,\,$        $\,\,\,$   &             & $\beta=1.90$, $a=0.0934$ fm  &        &           \\ \hline
$\,\,\,$  0.0080$\,\,\,$   &  0.162689   & $+$0.0275(4)     & 0.280(1)   & $24^3 \times 48$ & A \\  
$\,\,\,$        $\,\,\,$   &  0.163476   & $-$0.0273(2)     & 0.227(1)   \\ \hline              
$\,\,\,$  0.0080$\,\,\,$   &  0.162876   & $+$0.0398(1)     & 0.279(2)   & $24^3 \times 48$ & A,B \\  
$\,\,\,$        $\,\,\,$   &  0.163206   & $-$0.0390(1)     & 0.241(1)   \\ \hline              
$\,\,\,$        $\,\,\,$   &             & $\beta=1.95$, $a=0.0820$ fm  &         &          \\ \hline
$\,\,\,$  0.0020$\,\,\,$   &  0.160524   & $+$0.0363(1)     &            & $24^3 \times 48$ & A,B \\  
$\,\,\,$        $\,\,\,$   &  0.161585   & $-$0.0363(1)     &    \\ \hline                      
$\,\,\,$  0.0085$\,\,\,$   &  0.160826   & $+$0.0191(2)     & 0.277(2)   & $24^3 \times 48$ & A \\  
$\,\,\,$        $\,\,\,$   &  0.161229   & $-$0.0209(2)     & 0.259(1)   \\ \hline              
$\,\,\,$        $\,\,\,$   &             & $\beta=2.10$, $a=0.0644$ fm    &      &           \\ \hline
$\,\,\,$  0.0030$\,\,\,$   &  0.156042   & $+$0.0042(1)     & 0.127(2)   & $32^3 \times 64$ & B \\  
$\,\,\,$        $\,\,\,$   &  0.156157   & $-$0.0040(1)     & 0.129(3)   \\ \hline              
$\,\,\,$  0.0046$\,\,\,$   &  0.156017   & $+$0.0056(1)     & 0.150(2)   & $32^3 \times 64$ & A \\  
$\,\,\,$        $\,\,\,$   &  0.156209   & $-$0.0059(1)     & 0.160(4)   \\ \hline              
$\,\,\,$  0.0064$\,\,\,$   &  0.155983   & $+$0.0069(1)     & 0.171(1)   & $32^3 \times 64$ & A \\  
$\,\,\,$        $\,\,\,$   &  0.156250   & $-$0.0068(1)     & 0.180(4)   \\ \hline
\hline
\end{tabular}
\caption{Simulation parameters for the $N_f=4$
  ensembles~\cite{Alexandrou:2015sea} used in the calculation of the
  renormalization. The last column indicates which ensembles were used
  in Sections~\ref{sec:RI_renorm} (for the whole-operator
  renormalization) and \ref{sec:aux_renorm} (for the auxiliary-field
  renormalization).}
\label{Table:Z_ensembles}
\end{center}
\end{table}

\subsection{Whole operator approach and RI$'$-MOM scheme}
\label{sec:RI_renorm}

The Rome-Southampton approach~\cite{Martinelli:1994ty} and its
RI$^{(\prime)}$-MOM schemes are commonly used to determine
renormalization factors of local operators. Our prescription for the
nonlocal operator $\cO_\Gamma(z)$ closely follows
Refs.~\cite{Constantinou:2017sej, Alexandrou:2017huk} and the
improvements from Ref.~\cite{Alexandrou:2019lfo} for controlling
systematic uncertainties; we refer the reader to those references for
a more detailed discussion.

In Landau gauge and in momentum space, we compute the fermion
propagator $S_q$ [Eq.~\eqref{eq:quark_prop}] and the amputated vertex
function $\mathcal{V}_\cO$, with the operator $\cO$ inserted at zero
momentum transfer. We impose the conditions
\begin{gather}
\label{renorm}
\frac{Z^{\RI}_{\cal O}(z,\mu_0,m_\pi)}{Z^{\RI}_q(\mu_0,m_\pi)}\frac{1}{12} {\rm Tr} \left[{\cal V}_\cO(z,\vvp,m_\pi) \left({\cal V}_\cO^{\rm Born}(z,\vvp)\right)^{-1}\right] \Bigr|_{p^2{=}\mu_0^2} {=} 1\, ,\\[4ex]
Z^{\RI}_q(\mu_0,m_\pi)= \frac{1}{12} {\rm Tr} \left[(S_q(\vvp,m_\pi))^{-1}\, S_q^{\rm Born}(\vvp)\right] \Bigr|_{p^2=\mu_0^2}  \,, \quad
\end{gather}
at each value of $z$, where $X^\text{Born}$ is the tree-level value of
$X$. As a shorthand, we write $Z_V$ for the renormalization of the
unpolarized operator $\cO_{\gamma_0}$ and $Z_A$ for the helicity
operator $\cO_{\gamma_3\gamma_5}$. We choose the RI$'$ renormalization
scale, $\mu_0$, so that the vertex momentum $\vvp$ has the same
components in all spatial directions, that is,
$a \vvp = \frac{2\pi}{L_s}(n_t+\frac{1}{4},n,n,n)$ with integer $n$
and $n_t$. More precisely, we choose momenta with
$P_4 \equiv (\sum_i p_i^4) / (\sum_i p_i^2)^2 \le 0.32$, in order to
suppress finite-$a$ effects that break rotational
symmetry~\cite{Constantinou:2010gr, Alexandrou:2015sea}.

The renormalization factors are calculated on the $N_f=4$ ensembles
given in Table~\ref{Table:Z_ensembles}; these have the same bare
coupling $\beta$ as the $N_f=2+1+1$ ensembles used for the bare matrix
elements. The renormalization procedure can be summarized in the
following steps:
\begin{enumerate}
\item Calculation of $Z_{\cal O}$ for each ensemble of
  Table~\ref{Table:Z_ensembles}, and for several values of the
  renormalization scale $\mu_0$. We use $n_t \in [3,9]$, $n \in [2,4]$
  for the $24^3\times48$ ensembles, and $n_t \in [3,10]$,
  $n \in [3,5]$ for the $32^3\times64$ ensembles, and restrict to the
  momenta satisfying $P_4 \le 0.3$. The range of values for
  $(a\mu_0)^2$ is $[1,5]$ and $[1,4]$, for $24^3\times48$ and
  $32^3\times64$, respectively.
\item Averaging of the two ensembles at opposite
  $a\mu_{\rm PCAC}^{sea}$ values followed by chiral extrapolation of
  the form $Z_0 + (a \mu) Z_1$ (or quadratic in $a m_\pi$) for each
  lattice spacing. For $\beta=1.90$ we take the average of the four
  ensembles, as there is only one $a\mu$ value available. For all
  three $\beta$ values, we find a very mild dependence on the pion mass,
  similarly to what was found for other
  ensembles~\cite{Alexandrou:2019lfo}.
\item Conversion to the $\MSbar$ scheme and simultaneous evolution to
  the scale 2 GeV, using the expressions from
  Ref.~\cite{Constantinou:2017sej}.
\item Elimination of residual dependence on the RI$'$ scale by fitting
  to extrapolate $(a\mu_0)^2\to 0$. An extensive study on the choice
  of the renormalization scale and the corresponding systematic
  uncertainties can be found in Ref.~\cite{Alexandrou:2019lfo}. The
  optimal fit range for all $\beta$ values is
  $(a\mu_0)^2 \in [1,3]$.
\item Conversion to the $\MMS$ scheme, which is necessary in order to
  apply a matching formula that satisfies particle number
  conservation.
\end{enumerate}

\begin{figure*}
  \centering
  \includegraphics[width=0.495\textwidth]{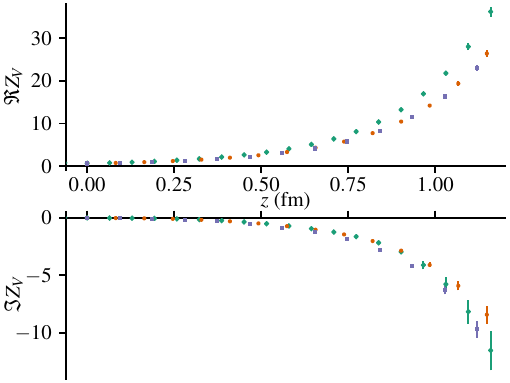}
  \includegraphics[width=0.495\textwidth]{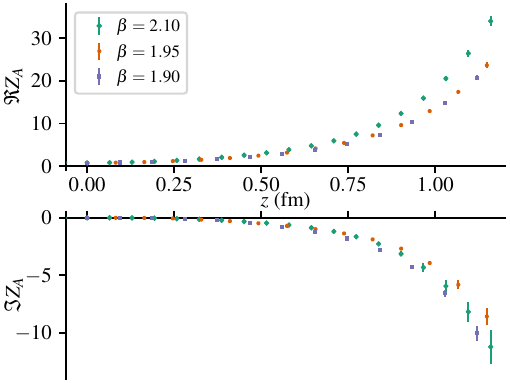}
  \caption{Real part (top) and imaginary part (bottom) of the
    renormalization factors for the unpolarized (left) and helicity
    (right) operators in the $\rm M\MSbar$ scheme at 2 GeV, determined
    using the whole-operator approach via the RI$'$-MOM intermediate
    scheme. Data at $\beta=2.10$, $\beta=1.95$ and $\beta=1.90$ are
    shown with green diamonds, orange circles, and blue squares,
    respectively.}
  \label{fig:Zfactors_MMS}
\end{figure*}

The final estimates for renormalization factors are shown in
Fig.~\ref{fig:Zfactors_MMS}. For the real part, the results with
$\beta=1.90$ and 1.95 are very similar, but the latter has a smaller
imaginary part. The finest lattice spacing, $\beta=2.10$, has a larger
real part. The renormalized matrix elements from the three lattice
spacings are shown in Fig.~\ref{fig:ME_RI_extrap} and their approach
to the continuum limit is discussed in Section~\ref{sec:continuum_me}.

\subsection{Auxiliary field approach and RI-xMOM scheme}
\label{sec:aux_renorm}

The auxiliary field approach~\cite{Craigie:1980qs, Dorn:1986dt,
  Ji:2017oey, Green:2017xeu} introduces a new field $\zeta(z)$ whose
propagator is a Wilson line along the $\vvzhat$ direction. This allows
the nonlocal operator in QCD to be represented using the local
operator $\phi\equiv\bar\zeta\psi$ in the extended theory:
\begin{equation}
  \cO_\Gamma(\vvx,z) = \left\langle \bar\phi(\vvx+z\vvzhat) \Gamma \tau_3 \phi(\vvx) \right\rangle_\zeta.
\end{equation}
The problem becomes that of renormalizing the action for $\zeta$ and
the composite operator $\phi$; one finds that three parameters are
sufficient to renormalize all operators
$\cO_\Gamma(z)$~\cite{Green:2017xeu}:
\begin{equation}
  \cO_\Gamma^R(z) = Z_\phi^2 e^{-m|z|} \left[
    \cO_{\Gamma}(z)
    + \sgn(z) r_\text{mix} \cO_{\{\gamma_z,\Gamma\}}
    + r_\text{mix}^2 \cO_{\gamma_z\Gamma\gamma_z}
  \right],
\end{equation}
where $m$ is linearly divergent, $Z_\phi$ is logarithmically
divergent, and $r_\text{mix}$ is finite and associated with chiral
symmetry breaking on the lattice. For our choices of $\Gamma$, the
anticommutator vanishes and the expression simplifies to
\begin{equation}
  \cO_\Gamma^R(z) = Z_\phi^2(1-r_\text{mix}^2)e^{-m|z|} \cO_\Gamma(z).
\end{equation}

We follow the approach in Refs.~\cite{Green:2017xeu, Green:2020xco} to
determine $m$ and $Z_\phi^2(1-r_\text{mix}^2)$, using the RI-xMOM
intermediate scheme and converting to $\MSbar$. Calculations are
performed using the most chiral $N_f=4$ twisted mass ensembles from
Ref.~\cite{Alexandrou:2015sea}, averaging over pairs of ensembles with
opposite PCAC masses rather than directly working at maximal twist. In
addition to the operator with stout-smeared links used for the bare
nucleon matrix elements, we also employ unsmeared links, which are
expected to have reduced discretization effects, in some intermediate
steps. After fixing to Landau gauge, we compute the position-space
$\zeta$ propagator
\begin{equation}
  S_\zeta(z)
  \equiv \langle \zeta(z\vvzhat)\bar\zeta(\vvzero)\rangle_{\text{QCD}+\zeta}
  = \langle W(z\vvzhat,\vvzero)\rangle_\text{QCD};
\end{equation}
the momentum-space quark propagator,
\begin{equation}\label{eq:quark_prop}
  S_q(\vvp) \equiv \int d^4\vvx e^{-i\vvp\cdot\vvx}
  \langle \chi(\vvx) \bar\chi(\vvzero) \rangle,
\end{equation}
where $\chi$ is a quark field in the twisted basis; and the
mixed-space Green's function for $\phi$,
\begin{equation}
  G_\phi(z,\vvp) \equiv \int d^4x e^{i\vvp\cdot\vvx}
  \langle \zeta(z\vvzhat) \phi(\vvzero) \bar\chi(\vvx) \rangle_{\text{QCD}+\zeta}.
\end{equation}
These renormalize as
\begin{align}
  S_\zeta^R(z) &= Z_\zeta e^{-m|z|} S_\zeta(z),\\
  S_q^R(\vvp) &= Z_q S_q(\vvp),\\
  G_\phi^R(z,\vvp) &= Z_\phi \sqrt{Z_\zeta Z_q} e^{-m|z|} G_\phi(z,\vvp).
\end{align}

\begin{figure}
  \centering
  \includegraphics[width=0.5\textwidth]{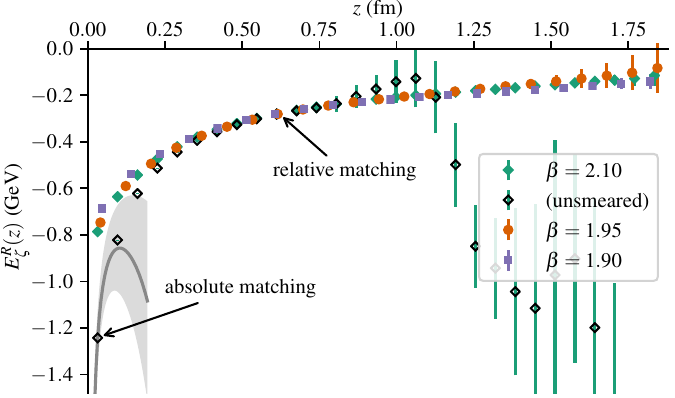}
  \caption{Renormalized $E_\zeta(z)$ versus $z$. Filled blue squares,
    orange circles, and green diamonds show the data with
    stout-smeared links on the coarse, medium, and fine lattice
    spacings, respectively. Diamonds with black outlines show the data
    on the fine lattice spacing without smearing. Note that a
    hypercubic rotation has been used to orient the Wilson line in the
    temporal direction to reduce finite-volume effects at large
    $z$. The curve shows the perturbative result based on the analytic
    three-loop calculation~\cite{Chetyrkin:2003vi, Melnikov:2000zc},
    the analytic partial four-loop
    calculations~\cite{Broadhurst:1994se, Grozin:2015kna,
      Grozin:2015kna, Grozin:2016ydd, Grozin:2017css,
      Marquard:2018rwx, Grozin:2018vdn, Bruser:2019auj}, and the
    numerical full four-loop calculation~\cite{Marquard:2018rwx}; its
    error band indicates the size of the $O(\alpha_s^4)$
    contribution.}
  \label{fig:Ezeta}
\end{figure}

To fix $m$, we evaluate the effective energy of the $\zeta$ propagator,
\begin{equation}
  E_\zeta(z) \equiv -\frac{d}{dz} \log \Tr S_\zeta(z),
\end{equation}
which is renormalized by adding $m$. We use the nearest-neighbor
lattice derivative. The relative matching among the three lattice
spacings is done at $z\approx 0.61$~fm. The absolute value of $m$ is
determined using unsmeared links on the finest lattice spacing, which
is expected to produce the smallest discretization effects, and
matching to the perturbative results for the static quark propagator
known to $O(\alpha_s^4)$~\cite{Green:2020xco, Chetyrkin:2003vi,
  Melnikov:2000zc, Broadhurst:1994se, Grozin:2015kna, Grozin:2015kna,
  Grozin:2016ydd, Grozin:2017css, Marquard:2018rwx, Grozin:2018vdn,
  Bruser:2019auj}. The results are shown in
Fig.~\ref{fig:Ezeta}. Except at short distance where discretization
effects are significant, the three lattice spacings are in good
agreement for the renormalized effective energy.

The other renormalization factors are determined using conditions
designed to eliminate dependence on $m$:
\begin{gather}
  \frac{-i}{12p^2Z_q^\RI} \Tr\left[ S_q^{-1}(\vvp) \slashed{\vvp} \right] = 1,\\
  \frac{Z_\zeta^\RI}{3} \frac{[\Tr S_\zeta(z)]^2}{\Tr S_\zeta(2z)} = 1,\\
  \frac{1}{12}
 \frac{Z_\phi^\RI(1\pm r_\text{mix})}{\sqrt{Z_\zeta^\RI Z_q^\RI}}
  \Re \Tr \left[ (1\pm\gamma_z) S_\zeta^{-1}(z) G_\phi(z,\vvp) S_q^{-1}(\vvp) \right] = 1.
\end{gather}
These are evaluated at the scale $\mu^2=p^2$, choosing
$\vvp=p_z\vvzhat$. This defines a family of renormalization schemes
that depend on the dimensionless quantity $y\equiv p_z z$. From the
above, we extract the relevant overall renormalization factor,
$Z_\phi^2(1-r_\text{mix}^2)$, at fixed kinematics. We then convert
$Z_\phi$ to the $\MSbar$ scheme using the one-loop expression from
Refs.~\cite{Green:2017xeu, Green:2020xco} and evolve to the scale
2~GeV using the two-loop anomalous dimension of the static-light
current~\cite{Ji:1991pr, Broadhurst:1991fz}.

\begin{figure}
  \centering
  \includegraphics[width=0.5\textwidth]{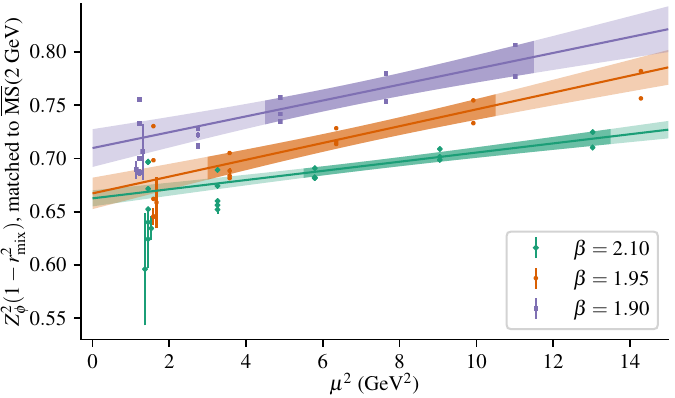}
  \caption{Renormalization factor $Z_\phi^2(1-r_\text{mix}^2)$
    determined using unsmeared links in the RI-xMOM scheme, matched to
    $\MSbar$, and evolved to scale 2~GeV. A hypercubic rotation has
    been used to orient the Wilson line and the quark momentum in the
    temporal direction. The multiple points at the same $\mu^2$ have
    different values of the RI-xMOM scheme parameter $y$. The lines with
    error bands give the extrapolation to $a^2\mu^2=0$ and the darker
    part of each error band indicates the fit range.}
  \label{fig:Z2}
\end{figure}

The determination of $Z_\phi^2(1-r_\text{mix}^2)$ is shown in
Fig.~\ref{fig:Z2}. As this is done at relatively high scales where the
perturbative matching and evolution are applicable, we do this using
unsmeared gauge links. Except at low $\mu^2$, the statistical
uncertainty is negligible compared with systematics. At each $\mu^2$,
we estimate the latter such that the spread of results for different
scheme parameters $y$ is covered. For each lattice spacing, we
extrapolate $a^2\mu^2$ to zero assuming a linear dependence; the
systematic uncertainty is propagated assuming a 50\% correlation
between every pair of points. Following the approach used in
Ref.~\cite{Green:2017xeu}, we match between unsmeared and smeared
links in the infrared regime at large $z$ and small $p^2$.

\begin{table}
  \centering
  \begin{tabular}{c|cc}
    $\beta$ & $am$ & $Z_\phi^2(1-r_\text{mix}^2)$ \\\hline
    1.90 & $-0.392(1)(57)$ & 0.986(25) \\
    1.95 & $-0.373(1)(50)$ & 0.908(20) \\
    2.10 & $-0.305(1)(37)$ & 0.907(11)
  \end{tabular}
  \caption{Renormalization parameters from the auxiliary-field
    approach, determined via the RI-xMOM intermediate scheme. The
    second uncertainty for the auxiliary field mass comes from the
    absolute matching onto perturbation theory and is fully correlated
    across the three ensembles.}
  \label{tab:RI-xMOM}
\end{table}

Final parameters for operators with stout-smeared links are given in
Table~\ref{tab:RI-xMOM}. The large uncertainty for the mass parameter
is caused by the absolute matching onto perturbation theory. At each
$z$, this absolute matching produces an overall factor applied to
$h_\Gamma(z)$ at all three lattice spacings. Therefore it can be
ignored when studying the approach to the continuum limit. However,
this uncertainty must be included when comparing continuum-limit
results against other renormalization approaches.

An additional perturbative conversion~\cite{Alexandrou:2019lfo} yields
results in the $\MMS$ scheme; this cancels a $\log(z^2)$ divergence in
the $\MSbar$-renormalized matrix element at short distance. However,
this conversion has only been computed at one-loop order, meaning that
the cancellation may be inexact and some part of the divergence may
still remain. The renormalized nucleon matrix elements for the three
lattice spacings are shown in Fig.~\ref{fig:ME_auxrenorm_extrap}.

\subsection{Ratio with zero-momentum matrix element}
\label{sec:p0ratio}

\begin{table}
  \centering
  \begin{tabular}{c|cc}
    Ensemble & $N_\text{conf}$ & $N_\text{samp}$ \\\hline
    A60 & 79 & 316 \\
    B55 & 54 & 216 \\
    D45 & 65 & 260
  \end{tabular}
  \caption{Statistics used for the nucleon matrix elements at zero
    momentum.}
  \label{tab:zero_momentum_stats}
\end{table}

The simplest way to cancel ultraviolet divergences is to compute
matrix elements of the same operator in different hadronic states and
then take their ratio. Here we choose to take the ratio of matrix
elements in a nucleon at nonzero momentum (i.e.\ those used throughout
this paper) with the same in a nucleon at rest,
\begin{equation}
  R_\Gamma(p_z,z) \equiv \frac{h_\Gamma(p_z,z;\mu)}{h_\Gamma(0,z;\mu)}.
\end{equation}
As the signal-to-noise problem is much milder in a nucleon at rest,
this requires a relatively inexpensive additional calculation: see
Table~\ref{tab:zero_momentum_stats}.

This ratio is similar to the reduced Ioffe-time distribution used in the
pseudo-PDF approach for parton
distributions~\cite{Radyushkin:2017cyf}. Although it is a different
observable than the $\MMS$-renormalized matrix elements used for
quasi-PDFs, it provides the opportunity to study the approach to the
continuum limit in a clean, controlled setting. As such, this section
can be seen as a preview of the continuum extrapolations of the
renormalized matrix element $h_\Gamma$ in
Section~\ref{sec:continuum_me}.

We consider variations of the continuum extrapolation in two different
ways. First: precisely which points should be used to obtain
$R_\Gamma(p_z,z)$ at zero lattice spacing? One option is to ignore
small differences in $p_z$ among the three ensembles, interpolate the
lattice data to a common value of $z$ in physical units, and then
perform the extrapolation. Alternatively, noting that parameter $x$ of
quasi-PDFs is Fourier-conjugate to the product $zp_z$, we can choose
to interpolate to a common value of $zp_z$ before extrapolating; this
could be more reliable because it accounts at leading order for the
small differences in $p_z$.

Second: what fit form should be used? As we have three lattice
spacings, we restrict ourselves to two-parameter fits. At $z=0$, the
operator $\cO$ is local, namely a vector or axial current; since we
work at maximal twist, this calculation benefits from automatic $O(a)$
improvement~\cite{Frezzotti:2003ni, Shindler:2007vp} and we
extrapolate using an affine function of $a^2$. For $z\neq 0$
(i.e.\ for the bulk of our data), $\cO$ is nonlocal and there can be
$O(a)$ contributions that are not eliminated by automatic
improvement~\cite{Green:2020xco}. In practice, it is not clear whether
we are in the regime where $O(a)$ contributions dominate; therefore,
we extrapolate using both affine functions of $a$ and of $a^2$.

\begin{figure*}
  \centering
  \includegraphics[width=0.495\textwidth]{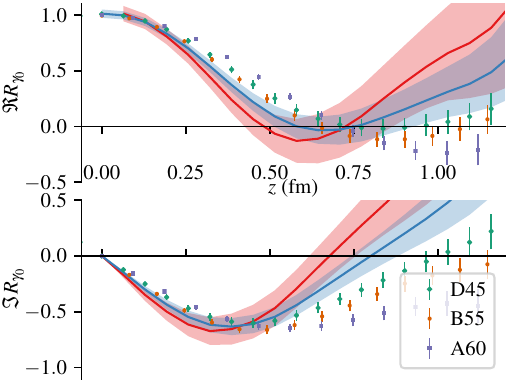}
  \includegraphics[width=0.495\textwidth]{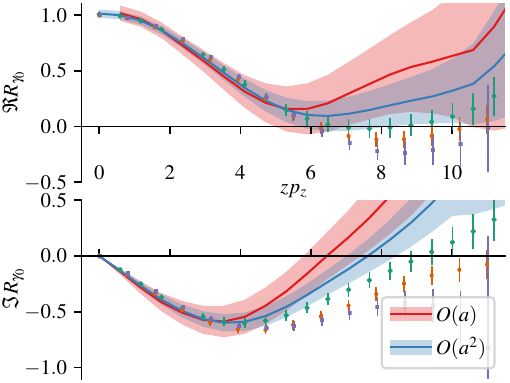}
  \caption{Real part (top) and imaginary part (bottom) of the ratio of
    unpolarized matrix elements $R_{\gamma_0}$ versus $z$ in physical
    units (left) and versus $zp_z$ (right). The curves with error
    bands depict the results from the continuum extrapolations
    assuming leading artifacts linear in $a$ (red) and quadratic in
    $a$ (blue).}
  \label{fig:ratio_unpol}
\end{figure*}

\begin{figure*}
  \centering
  \includegraphics[width=0.495\textwidth]{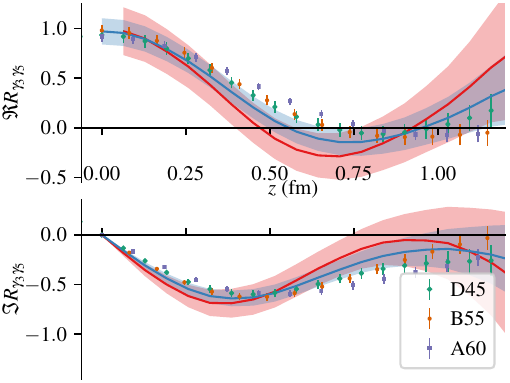}
  \includegraphics[width=0.495\textwidth]{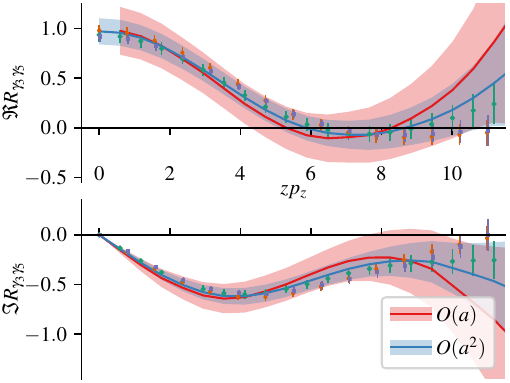}
  \caption{Ratio of helicity matrix elements
    $R_{\gamma_3\gamma_5}$. See the caption of
    Fig.~\ref{fig:ratio_unpol}.}
  \label{fig:ratio_hel}
\end{figure*}

The ratio data and their extrapolations are shown in
Figs.~\ref{fig:ratio_unpol} and \ref{fig:ratio_hel}. When plotted
versus $z$ in physical units, clear discrepancies between the three
ensembles are visible and for most of the parameter space, the lattice
data from the coarsest ensemble are more than one standard deviation
away from the extrapolations. These discrepancies are reduced when
plotting the data versus $zp_z$, although they remain significant for
the unpolarized data at large $z$.

From this study, it appears that performing the extrapolation at fixed
values of $zp_z$ is a better approach. For the unpolarized matrix
elements with $zp_z < 5$ and for the helicity matrix elements, lattice
artifacts have a modest effect and are well under control, with the
$O(a)$ and $O(a^2)$ extrapolations in good agreement. For the
unpolarized matrix elements with $zp_z > 5$, there is a stronger
dependence on $a$ and worse agreement between the two extrapolations;
this suggests that at longer distances the lattice artifacts are less
well controlled.

\section{Continuum limit}
\label{sec:continuum}

\subsection{Renormalized matrix elements}
\label{sec:continuum_me}

Based on our study of the ratios of matrix elements in the previous
section, we choose to linearly interpolate our $\MMS$-scheme
renormalized matrix elements to common values of $zp_z$ and then
perform continuum extrapolations at each interpolated point. We again
extrapolate in two ways, assuming lattice artifacts are either linear
or quadratic in the lattice spacing.

\begin{figure*}
  \centering
  \includegraphics[width=0.495\textwidth]{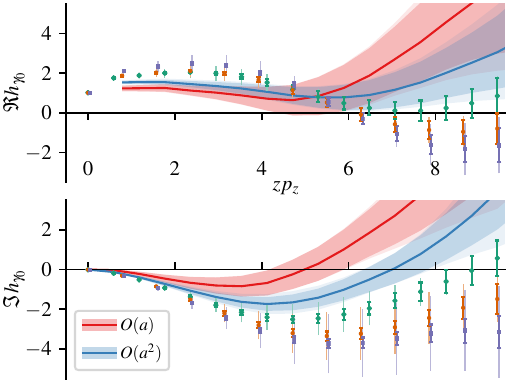}
  \includegraphics[width=0.495\textwidth]{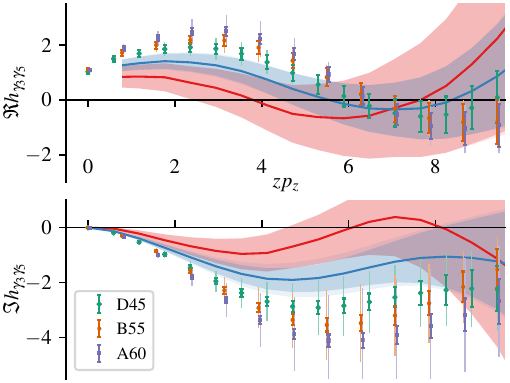}
  \caption{Matrix elements renormalized using the auxiliary field
    approach via matching from the intermediate RI-xMOM scheme: real
    part (top) and imaginary part (bottom) of the unpolarized (left)
    and helicity (right) matrix elements, converted to the $\MMS$
    scheme at scale 2~GeV. The curves with error bands depict the
    results from the continuum extrapolations assuming leading
    artifacts linear in $a$ (red) and quadratic in $a$ (blue). The
    outer error bars (without endcaps) and outer error bands include
    the uncertainty from the absolute matching of the auxiliary field
    mass onto perturbation theory, which is fully correlated among the
    three ensembles.}
  \label{fig:ME_auxrenorm_extrap}
\end{figure*}

Figure~\ref{fig:ME_auxrenorm_extrap} shows the matrix elements
renormalized using the auxiliary field approach and their continuum
extrapolations. For most values of $zp_z$, there is a large dependence
on the lattice spacing and the extrapolated values are far from those
of the individual ensembles. The extrapolations tend to reduce the
magnitude of the matrix element, except for the unpolarized case at
large $zp_z$, where both the real and imaginary parts are positive and
growing. For the unpolarized matrix element, the $O(a)$ and $O(a^2)$
extrapolations are not in good agreement, particularly in the real
part at small $zp_z$ and the imaginary part at medium $zp_z$.

\begin{figure*}
  \centering
  \includegraphics[width=0.495\textwidth]{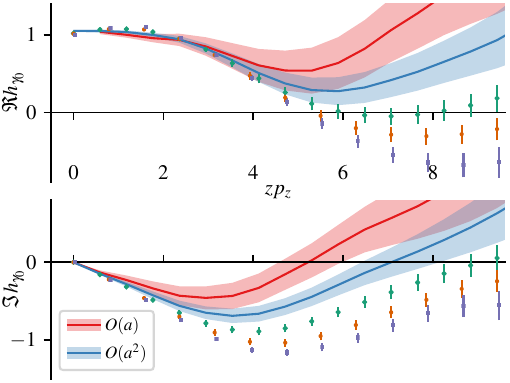}
  \includegraphics[width=0.495\textwidth]{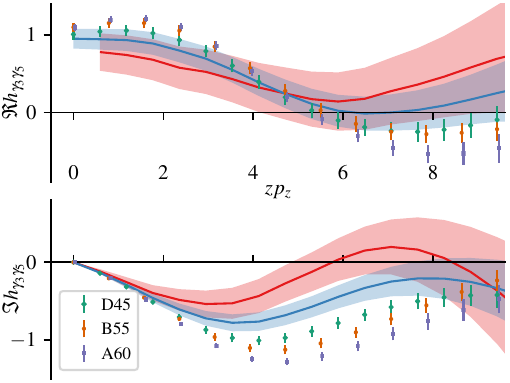}
  \caption{Matrix elements renormalized using the whole operator
    approach via matching from the intermediate RI$'$-MOM scheme: real
    part (top) and imaginary part (bottom) of the unpolarized (left)
    and helicity (right) matrix elements, converted to the $\MMS$
    scheme at scale 2~GeV. The curves with error bands depict the
    results from the continuum extrapolations assuming leading
    artifacts linear in $a$ (red) and quadratic in $a$ (blue).}
  \label{fig:ME_RI_extrap}
\end{figure*}

\begin{figure*}
  \centering
  \includegraphics[width=\textwidth]{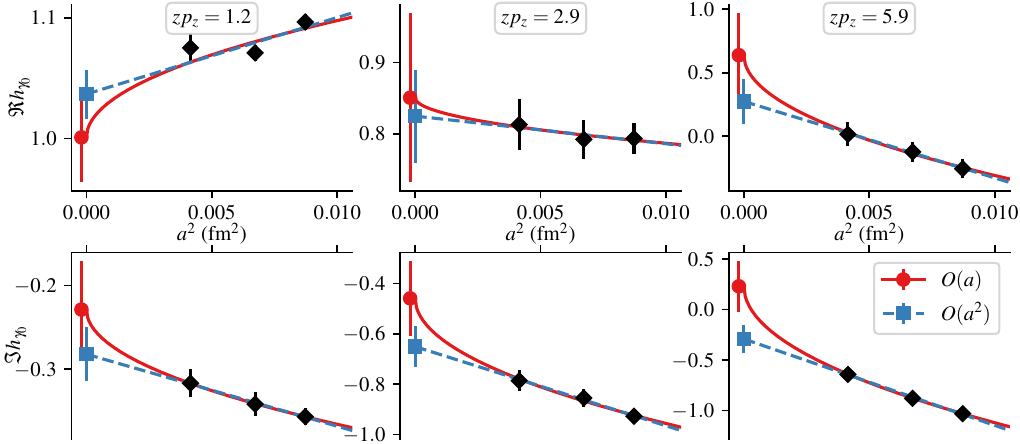}
  \caption{Continuum extrapolation of unpolarized matrix elements
    renormalized using the whole operator approach: real part (top)
    and imaginary part (bottom), versus $a^2$. Results are shown for
    $zp_z=1.2$ (left), 2.9 (center), and 5.9 (right). Black diamonds
    show the data from ensemble D45 and interpolated data from
    ensembles B55 and A60. The solid red curve and red circle show the
    $O(a)$ extrapolation; the dashed blue line and blue square show
    the $O(a^2)$ extrapolation.}
  \label{fig:extrap_unpol}
\end{figure*}

\begin{figure*}
  \centering
  \includegraphics[width=\textwidth]{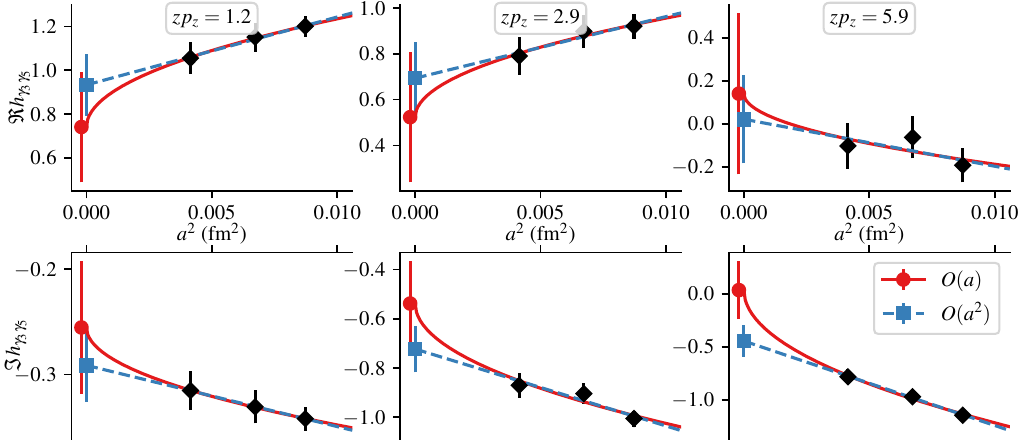}
  \caption{Continuum extrapolation of helicity matrix elements
    renormalized using the whole operator approach. See the caption of
    Fig.~\ref{fig:extrap_unpol}.}
  \label{fig:extrap_hel}
\end{figure*}

Matrix elements renormalized using the whole-operator approach are
shown in Fig.~\ref{fig:ME_RI_extrap}, along with their continuum
extrapolations. Qualitatively, the picture is similar to the
auxiliary-field renormalization approach, except that at small $zp_z$,
the real part of the matrix elements from the three lattice spacings
are in better agreement, producing a milder effect from the continuum
extrapolation and a better agreement between the two
extrapolations. The latter is especially true for the unpolarized
matrix element. Details of these continuum extrapolations for selected
values of $zp_z$ are shown in Figs.~\ref{fig:extrap_unpol} and
\ref{fig:extrap_hel}. Clearly, our lever arm in $a$ is limited, which
makes it difficult to detect a preference for either of the two fits;
this also produces a large uncertainty for the $O(a)$ extrapolations.

\begin{figure*}
  \centering
  \includegraphics[width=0.495\textwidth]{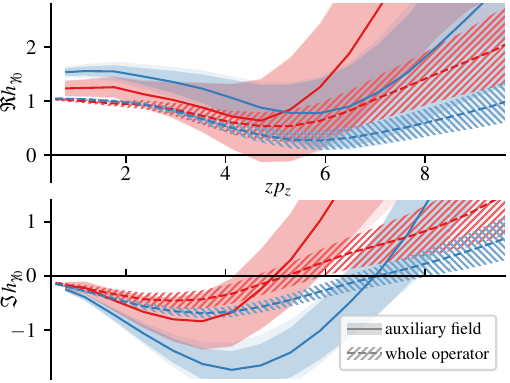}
  \includegraphics[width=0.495\textwidth]{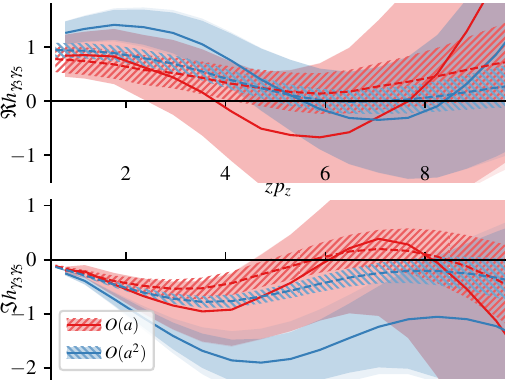}
  \caption{Comparison of continuum-extrapolated matrix elements: real
    part (top) and imaginary part (bottom) of the unpolarized (left)
    and helicity (right) matrix elements, in the $\MMS$ scheme at
    scale 2~GeV. Solid bands show data using the auxiliary field
    approach via the RI-xMOM intermediate scheme and hatched bands
    show results using the whole-operator approach via the RI$'$-MOM
    intermediate scheme.}
  \label{fig:ME_both_extrap}
\end{figure*}

Results from the two renormalization approaches are compared in
Fig.~\ref{fig:ME_both_extrap}. The whole operator approach tends to
produce a smaller central value and a smaller uncertainty than the
auxiliary field method. For the imaginary part of the matrix elements,
the $O(a^2)$ auxiliary-field extrapolation is in significant
disagreement with both of the whole-operator extrapolations. In
contrast, the $O(a)$ result using the auxiliary field method is
largely compatible with both whole-operator extrapolations, for low to
medium values of $zp_z$. This suggests that there may be significant
$O(a)$ lattice artifacts in the determination of the auxiliary field
renormalization parameters and that it is necessary to account for
them when taking the continuum limit.

\begin{figure*}
  \centering
  \includegraphics[width=0.495\textwidth]{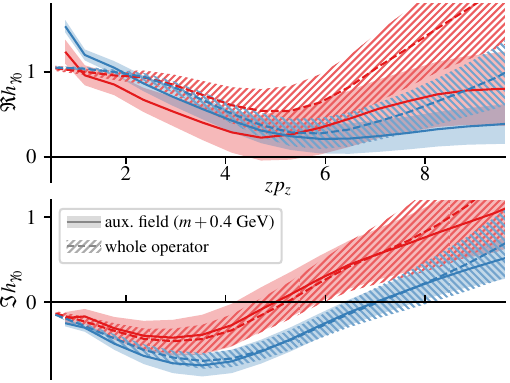}
  \includegraphics[width=0.495\textwidth]{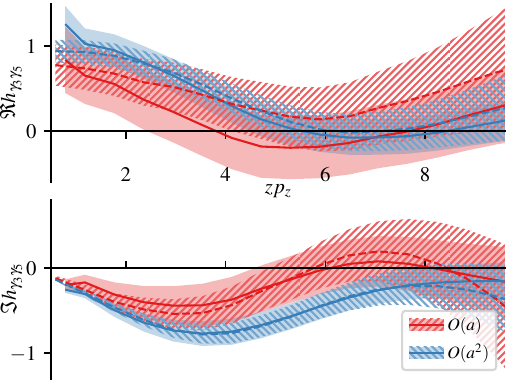}
  \caption{Comparison of continuum-extrapolated matrix elements, after
    reducing the magnitude of the auxiliary-field mass renormalization
    by 0.4~GeV. See the caption of Fig.~\ref{fig:ME_both_extrap}.}
  \label{fig:ME_both_adj}
\end{figure*}

Since renormalization in the auxiliary field approach is determined by
just two parameters, one might ask whether there exist parameters that
produce results compatible with the whole operator
method. Figure~\ref{fig:ME_both_adj} shows the effect of reducing the
magnitude of the auxiliary-field mass renormalization parameter by
$\delta m=0.4$~GeV. Although this adjustment is hard to justify from
the analysis in Section~\ref{sec:aux_renorm}, in
Ref.~\cite{Ji:2020brr} it was shown that its effect on quasi-PDFs is
suppressed by the factor $\delta m/p_z$ at large momentum. This change
produces good agreement for the imaginary part of the matrix
elements. However, some discrepancies remain for the real part,
particularly in the unpolarized case at small $zp_z$, where the slope
of the auxiliary-field result is considerably steeper than the
whole-operator data.

In the rest of this paper where we examine the effect on parton
distributions, we will focus on the more precise data renormalized
using the whole-operator method. However, we will continue to compare
$O(a)$ and $O(a^2)$ extrapolations since they are not in complete
agreement and we have no a priori reason to prefer one over the other.

\subsubsection{Comparison with phenomenology}

\begin{figure*}
  \centering
  \includegraphics[width=0.495\textwidth]{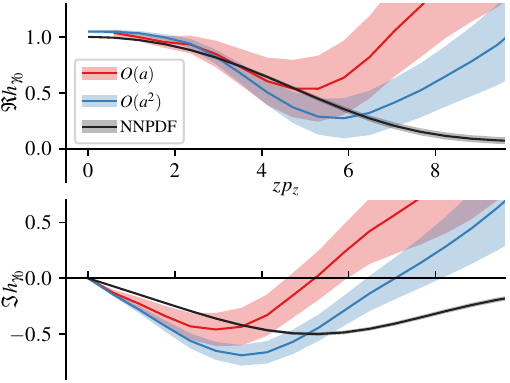}
  \includegraphics[width=0.495\textwidth]{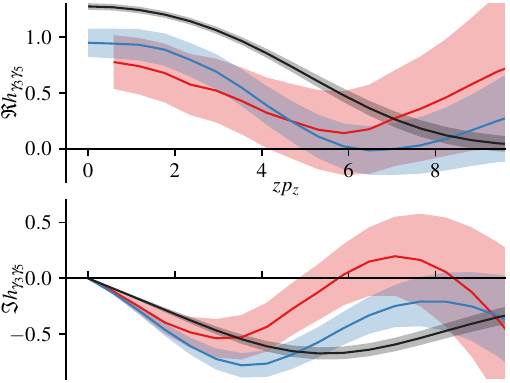}
  \caption{Unpolarized (left) and helicity (right) matrix elements
    from continuum extrapolation of lattice data renormalized using
    the whole operator approach via the RI$'$-MOM intermediate scheme
    (blue, red) and from the inverse Fourier transform of the
    quasi-PDFs obtained by applying inverse matching to phenomenological PDFs
    from NNPDF~\cite{Ball:2017nwa,Nocera:2014gqa} (dark gray). Note
    that in the lattice calculation, the pion mass is much larger than
    in nature, so that exact agreement should not be expected.}
  \label{fig:pheno_pz}
\end{figure*}

Before transforming the position-space matrix elements to obtain PDFs
and comparing directly with phenomenology, we perform the reverse
exercise. Starting with phenomenological parton distributions
determined by NNPDF~\cite{Ball:2017nwa,Nocera:2014gqa}, we invert the
matching and the Fourier transform to determine the position-space matrix
elements that yield those PDFs, up to higher-order corrections in the
matching. Figure~\ref{fig:pheno_pz} compares this with the
continuum-extrapolated lattice matrix elements. Full agreement cannot
be expected, since the lattice calculation was done at a heavy pion
mass and other systematics such as the dependence on $p_z$ and
finite-volume effects have not been included in this study.

The real part of the unpolarized matrix elements show reasonable
agreement for $zp_z<5$; in the same range, the helicity matrix
elements from the lattice lie below those from phenomenology. The
helicity case can be partly understood by recalling that at heavy pion
masses, the nucleon axial charge (i.e.\ the helicity matrix element at
$z=0$) lies below its physical value. At short distances, the
imaginary parts of the lattice data have larger (more negative) slopes
than phenomenology; the $O(a)$ extrapolations are consistent with the
latter at the $1\sigma$ level whereas the $O(a^2)$ extrapolations are
not. At nonzero lattice spacing, the slope is even larger and in worse
agreement with NNPDF, so that the continuum extrapolation produces
results that lie closer to phenomenology.

At larger values of $zp_z$, there is a qualitative difference: the
phenomenological curves tend steadily toward zero, whereas the lattice
data do not. This is especially true for the unpolarized lattice matrix
elements, of which both the real and imaginary parts are positive and
increasing at large distances. At the coarsest lattice spacing, the
lattice data lie well below zero (see Fig.~\ref{fig:ME_RI_extrap}), so
it appears that the continuum extrapolation may be an
overcorrection. Another way to characterize the imaginary part is via
the position of the minimum of the curve: in the lattice data, it lies
at a shorter distance than in phenomenology. This is consistent with
the general expectation that correlation functions are shorter ranged
at heavier pion masses.

\subsection{Parton distributions}

In this section, we present the main results of this paper, namely the
effect of the continuum extrapolation on PDFs. However, we first
discuss another source of systematic uncertainty: how to perform the
Fourier transform in the definition of the quasi-PDF using a finite
set of position-space data. We illustrate this using data on the
finest ensemble, D45. Next, we perform the continuum extrapolation at
fixed $x$, using the PDFs determined on each ensemble, and compare the
result with the PDF determined from the continuum-limit matrix
elements obtained in the previous section. Finally, we compare our
continuum-limit PDFs with phenomenology.

\subsubsection{Reconstruction techniques}

As given in Eq.~\eqref{eq:quasi-PDF}, the quasi-PDF $\tilde q(x)$ is
obtained from a Fourier transform (FT) of the renormalized matrix
elements $h(z)$. In practice, we obtain $h(z)$ at intervals of the
lattice spacing\footnote{When analyzing the continuum-limit $h(z)$ we
  sample it at intervals of the finest lattice spacing, which we
  simply denote $a$ in this context.}, i.e.\ $z/a\in\mathbb{Z}$. It is
also necessary to truncate the FT at $|z|\leq \zmax$, both because of
the finite lattice size, which imposes $\zmax\lesssim L/2$, and
because of growing statistical uncertainty at large $|z|$. Together,
these have the effect of replacing the continuous FT by a truncated
discrete FT (DFT):
\begin{align}
  \frac{p_z}{2\pi} \int_{-\infty}^{\infty}dz\,e^{-ix p_z z}\rightarrow \frac{p_z}{2\pi} a \sum_{z/a=-\zmax/a}^{\zmax/a} e^{-ix p_z z}.
  \label{eq:standard_FT}
\end{align}

\begin{figure*}
    \centering
    \includegraphics{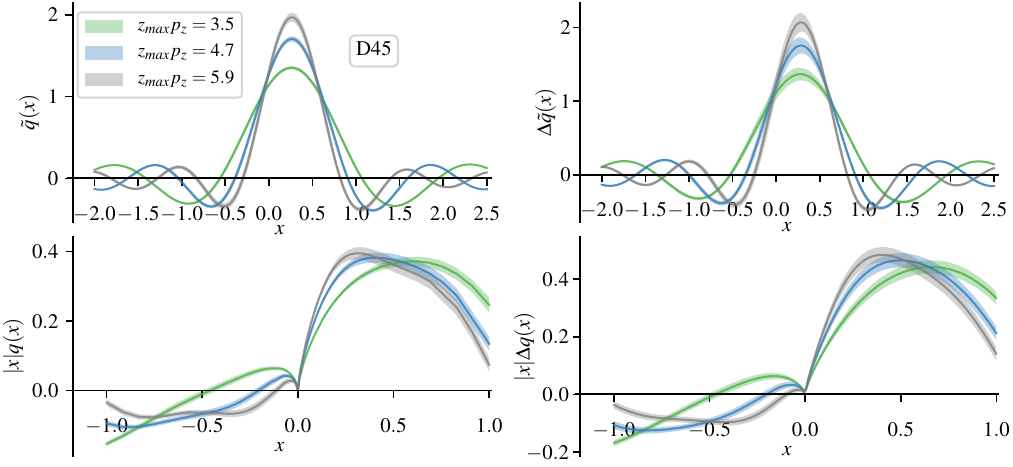}
    \caption{Unpolarized (left) and  helicity (right) quasi-PDFs (top panels) and PDFs (bottom panels) of the D45 ensemble for different values of the cutoff $\zmax$. The curves with cutoff at $\zmax p_z=\{ 3.5, 4.7, 5.9 \}$ are depicted in green, blue and gray.}
    \label{fig:cutoff_dependence}
\end{figure*}

The discrete sampling makes the result formally periodic, so that it
must be cut off at $|x|\leq \pi/(ap_z)$, which is at least 4 in our
setup.  The truncation introduces an additional systematic
uncertainty~\cite{Alexandrou:2019lfo}, as shown using ensemble D45 in
Fig.~\ref{fig:cutoff_dependence} for quasi-PDFs and PDFs. The latter
are obtained by applying the matching procedure and nucleon mass
corrections~\cite{Chen:2016utp}. For the quasi-PDF, the effect of
truncation is that one obtains a convolution of the desired result:
\begin{equation}
  \begin{aligned}
  \tilde q^\text{DFT}(x) &= \frac{ap_z}{2\pi} \int_{-\pi/(ap_z)}^{\pi/(ap_z)} dx'\,
  \frac{ \sin((x-x') p_z (\zmax + a/2)) }{ \sin((x-x') p_z a/2) } \tilde q(x')\\
  & \xlongrightarrow{a\to 0}
  \frac{1}{\pi} \int_{-\infty}^\infty dx'\, \frac{ \sin((x-x') \zmax p_z) }{(x-x')} \tilde q(x'),
  \end{aligned}
\end{equation}
so that any features narrower in $x$ than $(\zmax p_z)^{-1}$ are
smeared out. This is clearly visible in
Fig.~\ref{fig:cutoff_dependence}, where smaller values of $\zmax p_z$
are associated with broader quasi-distributions. The effect is reduced
after applying the matching to obtain PDFs: results with $\zmax
p_z=4.7$ and 5.9 are very similar. However, for $\zmax p_z=3.5$, both
the unpolarized and helicity PDFs have qualitatively quite different
behaviour, with a higher value for $x$ between roughly $-0.7$ and
$-0.1$ and a larger slope for $x$ less than $-0.3$ as well as a
smaller slope at small positive $x$ and a peak at larger $x$ in the
positive region.

Since the Fourier transform introduces a systematic uncertainty, we
supplement the naïve truncated FT with more sophisticated reconstruction
techniques~\cite{Karpie:2019eiq, Alexandrou:2020tqq}. In these
approaches, obtaining the Fourier transform from a finite number of
data points is seen as an ill-defined inverse problem. Its solution is
not unique and one approach is to use explicit models for the shape of
the (quasi-)PDF. By contrast, we choose to use two approaches that do
not contain an explicit model: the Backus-Gilbert method, first
applied for PDFs calculations in~\cite{Karpie:2019eiq} and the
Bayes-Gauss-Fourier Transform (BGFT)~\cite{Alexandrou:2020tqq}. These
two procedures address the reconstruction problem as follows.
\begin{description}
\item[Backus-Gilbert (BG)] The inverse problem is obtained by
  inverting Eq.~\eqref{eq:quasi-PDF} to write the real and imaginary
  parts of the unpolarized matrix element in terms of the quasi-PDF:
  \begin{equation}\label{eq:inverse_problem}
    \begin{aligned}
      \Re h_{\gamma_0}(p_z,z;\mu) &= \int_0^\infty dx\, \cos(x p_z z) \, \tilde q_+(x,p_z;\mu),\\
      \Im h_{\gamma_0}(p_z,z;\mu) &= \int_0^\infty dx\, \sin(x p_z z) \, \tilde q_-(x,p_z;\mu),\\
    \end{aligned}
  \end{equation}
  where for $x\geq 0$,
  $\tilde q_\pm(x) = \tilde q(x) \pm \tilde q(-x)$, and likewise for
  the helicity case\footnote{For the unpolarized case, this is not the
    same as the convention commonly used for PDFs, where
    $q_\pm(x) \equiv q(x)\pm\bar q(x) = q(x)\mp q(-x)$. For helicity,
    $\Delta q_\pm(x) \equiv \Delta q(x)\pm\Delta\bar q(x) = \Delta
    q(x)\pm\Delta q(-x)$.}. The reconstruction is applied
  independently to $\tilde q_+$ and $\tilde q_-$, so for brevity we
  describe the procedure applied to $\tilde q_+$. We also omit the
  labels $p_z$ and $\mu$. For each $x$, the solution is assumed to be
  a linear combination of the finite set of lattice data:
  \begin{equation}
    \tilde q_+^\text{BG}(x) = \sum_{z/a=0}^{\zmax/a} \textbf{a}_+(x,z) \Re h_{\gamma_0}(z),
  \end{equation}
  where $\textbf{a}_+$ can be understood as an approximation to the
  inverse of the Fourier transform in
  Eq.~\eqref{eq:inverse_problem}. The accuracy of this approximation
  is governed by the function
  \begin{equation}
    \Delta_+(x,x') = \sum_{z/a=0}^{\zmax/a} \textbf{a}_+(x,z) \cos(x' p_z z)
  \end{equation}
  that approximates $\delta(x-x')$. Specifically, the result is an
  integral over the quasi-PDF:
  \begin{equation}
    \tilde q_+^\text{BG}(x) = \int_0^\infty dx'\, \Delta_+(x,x') \tilde q_+(x').
  \end{equation}
  The function $\textbf{a}_+$ is determined by the Backus-Gilbert
  procedure~\cite{BackusGilbert}, which minimizes the width of
  $\Delta_+(x,x')$. For more details, see
  Refs.~\cite{Karpie:2019eiq, Bhat:2020ktg}.

\item[Bayes-Gauss-Fourier Transform (BGFT)] Rather than directly
  reconstructing $\tilde q$ from the lattice data, this procedure
  reconstructs a continuous form of the position-space matrix elements
  for all values of $z$:
  \begin{equation}
    h(z),\, z/a\in\{0,\pm1,\pm2,\dots,\pm\zmax/a\} \longrightarrow
    h^\text{GPR}(z),\, z\in\mathbb{R}.
  \end{equation}
  For this, we apply a nonparametric regression technique, based on
  Bayesian inference, called Gaussian process regression
  (GPR)~\cite{williams2006gaussian}. This allows us to incorporate
  into the prior distribution the asymptotic behavior of the matrix
  elements (expected to decay to zero), as well as their smoothness
  properties. The result is continuous, defined for all real $z$, and
  has a Fourier transform computable in closed form. Taking the FT of
  $h^\text{GPR}(z)$, we refer to the result as
  $\tilde q^\text{BGFT}(x)$. More details are given in
  Ref.~\cite{Alexandrou:2020tqq}.
\end{description}

\begin{figure*}
    \centering
    \includegraphics{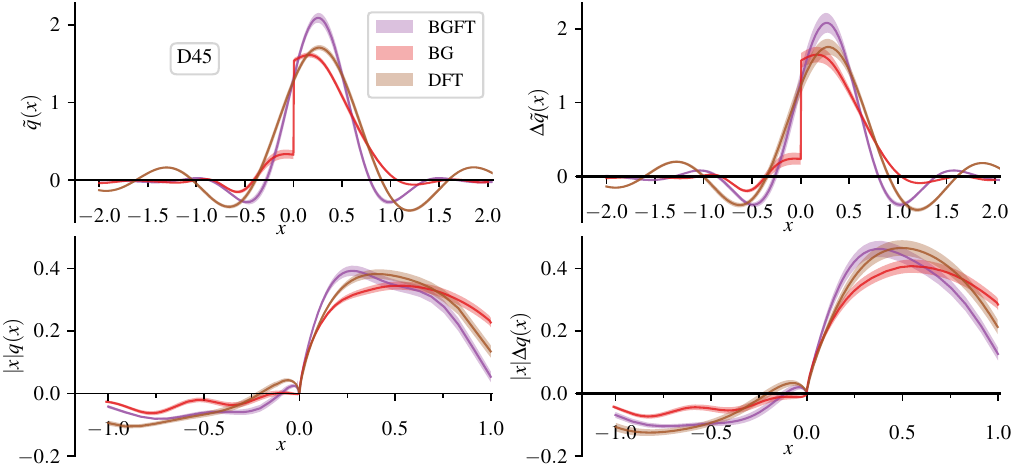}
    \caption{Comparison of quasi-PDFs (top panels) and PDFs (bottom panels) of the D45 ensemble obtained from Bayes-Gauss-Fourier Transform (BGFT), Backus-Gilbert (BG) and discrete FT (DFT) for the unpolarized (left) and helicity (right).}
    \label{fig:comparison_BG_BGFT_DFT}
\end{figure*}

In Fig.~\ref{fig:comparison_BG_BGFT_DFT}, we compare results from
the truncated discrete Fourier transform, Eq.~\eqref{eq:standard_FT},
and the BG and BGFT reconstruction methods described above, again using
ensemble D45 as our reference data set. For a fair comparison, in all
cases we use $\zmax p_z=4.7$. We begin by discussing the quasi-PDFs
(upper two panels). The most striking
difference is that the Backus-Gilbert result has a discontinuity at
$x=0$ that is not present in the other results. This is because
$\tilde q_-^\text{BG}(x)$ is not constrained to vanish at $x=0$. Such
a discontinuity could occur if $\Im h(z)$ has a slowly decaying tail
$\sim 1/z$. For $x$ between $-0.5$ and $1.0$, the DFT and BGFT results
are similar, although the BGFT distribution is slightly
narrower. For larger values of $|x|$, the DFT produces stronger
oscillations, which are suppressed by the BGFT. The BG result is the
outlier, being considerably smaller at small negative $x$ and also
having a smaller dip below zero.

We next discuss the physically relevant parton distributions,
obtained after matching and nucleon mass corrections (lower two panels).
For most values of $x$, the DFT and BGFT method produce very similar
results, although for BGFT the the dip below zero in the antiquark region
occurs at smaller negative $x$ and the magnitude is smaller at $x=-1$ and $+1$.
Again, the BG result is somewhat different: in the
antiquark region at small negative $x$, the small positive bump is gone
and the result is either consistent with zero (unpolarized) or slightly
negative (helicity). This
discrepancy at small $x$ may be associated with a lack of data for the
matrix element at large $|z|$; better data or a more rigorous
understanding of the large-$|z|$ behavior could help to improve this
situation. In the quark region for $x$ greater than about 0.5, the
BG result has a much weaker downward trend than the other two methods.
Given that the DFT produces a result not
substantially different from BGFT, we exclude the DFT from further
analyses presented in the next sections.

\subsubsection{Continuum extrapolation}

In what follows, we compare the distributions at finite lattice
spacings with continuum extrapolations. In the reconstruction of the
quasi-PDFs we use the lattice data with $|zp_z|\leq \zmax p_z=4.7$, at
which point either the real part or the imaginary part of the
continuum matrix element is compatible with zero, as shown in
Fig.~\ref{fig:ME_RI_extrap}. Moreover, we estimate the systematic
uncertainty from this choice of the cutoff by varying $\zmax$:
\begin{equation}
\varepsilon _{\rm cutoff}(x)=\frac{\vert   q_{\zmax p_z=5.9}(x)-q_{\zmax p_z=3.5}(x)\vert}{2}.
\end{equation}
Finally, we estimate the combined uncertainty as the quadrature sum of
$\varepsilon _{\rm cutoff}(x)$ and the statistical uncertainty.

\begin{figure*}
  \centering
 \includegraphics[width=\textwidth]{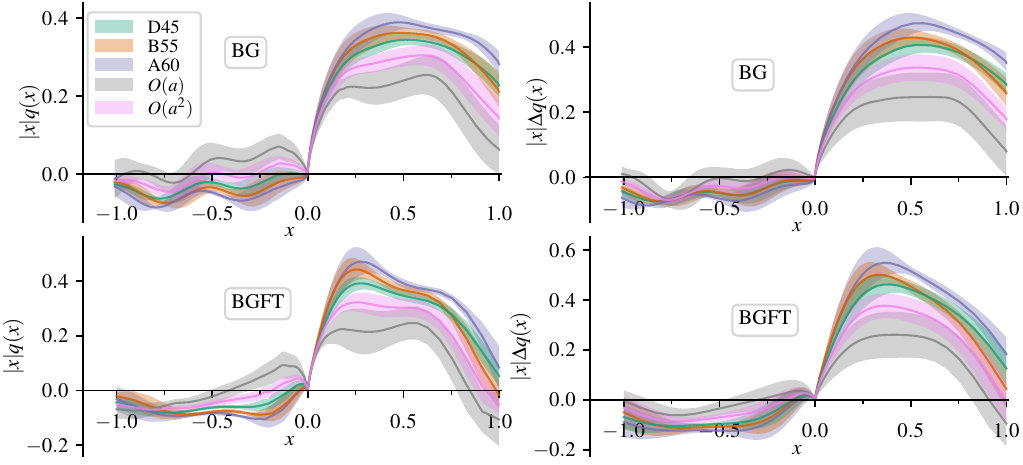}
 \caption{Matched unpolarized (left) and helicity (right) PDFs
   obtained using the gauge ensembles A60 (blue), B55 (orange), D45
   (green), whose lattice spacings are reported in
   Table~\ref{tab:ensembles}. The PDF in the continuum, after $O(a)$
   extrapolation (gray) and $O(a^2)$ extrapolation (pink), is also
   shown.}
  \label{fig:continuum_limit_3ensembles}
\end{figure*}

One approach for obtaining continuum-limit PDFs is to take the PDF
determined on each ensemble and then perform an $O(a)$ or $O(a^2)$
extrapolation of the data at each $x$. This is shown in
Fig.~\ref{fig:continuum_limit_3ensembles}, for both unpolarized and
helicity PDFs determined using the BG and BGFT methods. In the quark
region with $x$ between roughly 0 and 0.7, the PDFs decrease
monotonically with the lattice spacing; at larger $x$, the D45 data
(with the finest lattice spacing) move relatively upward to lie
between those of the other two ensembles. For all $x>0$, the $O(a^2)$
extrapolation lies below all of the individual lattice spacings and
the $O(a)$ extrapolation is even lower. Using the BGFT approach, both
of the extrapolations are consistent with the expected value of zero
at $x=1$, whereas for BG, this is true only of the $O(a)$
extrapolation. In the antiquark region, the extrapolated results lie
above the PDFs determined at finite lattice spacing, except for the
BGFT unpolarized distribution near $x=-1$. This produces a more
prominent positive region at small negative $x$, particularly in the
unpolarized case. At larger negative $x$, the extrapolations are
generally closer to zero.

\begin{figure*}
    \centering
    \includegraphics[width=\linewidth]{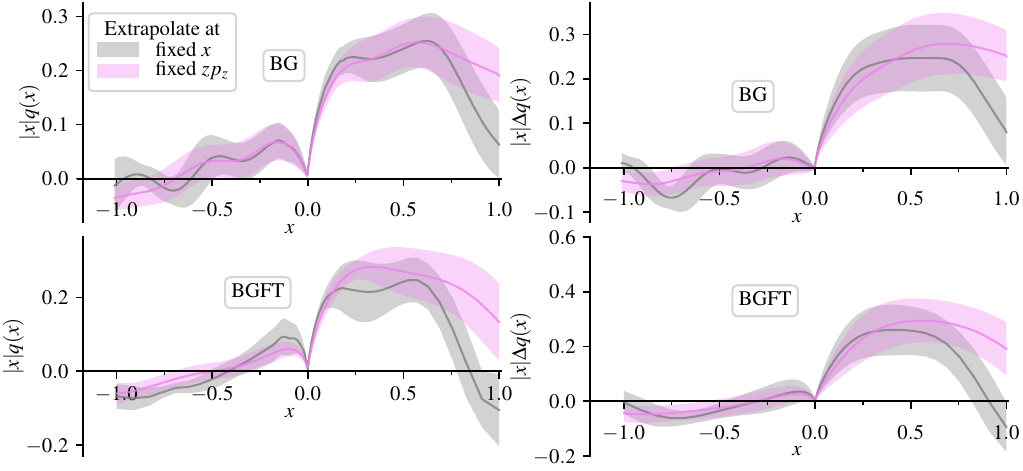}
    \caption{Comparison between the results for the unpolarized (left)
      and helicity (right) PDFs in the continuum limit obtained with
      the $O(a)$ extrapolation at fixed $x$ (gray; see
      Fig.~\ref{fig:continuum_limit_3ensembles}) and at fixed $zp_z$
      (pink, based on the continuum-limit data in
      Fig.~\ref{fig:ME_RI_extrap}). The distributions has been
      obtained using the BG (top panels) and BGFT (bottom panels)
      reconstruction techniques.}
    \label{fig:comparison_option_1_2}
\end{figure*}

Another approach is to obtain PDFs from the continuum limit of $h(z)$
as determined in Section~\ref{sec:continuum_me} by extrapolating data
at fixed $zp_z$. By changing the order in which the continuum limit
and the combination of the Fourier transform and PDF matching are
performed, we obtain results affected by different systematic
effects. The comparison of the $O(a)$ extrapolations from both
approaches is shown in Fig.~\ref{fig:comparison_option_1_2}. They are
consistent within uncertainties, except near $x=1$, where the
fixed-$x$ extrapolation is in all cases lower than the fixed-$zp_z$
extrapolation and only the former is consistent with zero at $x=1$.

For comparing with phenomenology in the next section, we take the
fixed-$x$ extrapolation as our central value and add an additional
systematic uncertainty in quadrature, namely half the difference with the
fixed-$zp_z$ extrapolation.

\subsubsection{Comparison with phenomenology}

\begin{figure*}
  \centering
    \includegraphics[width=\textwidth]{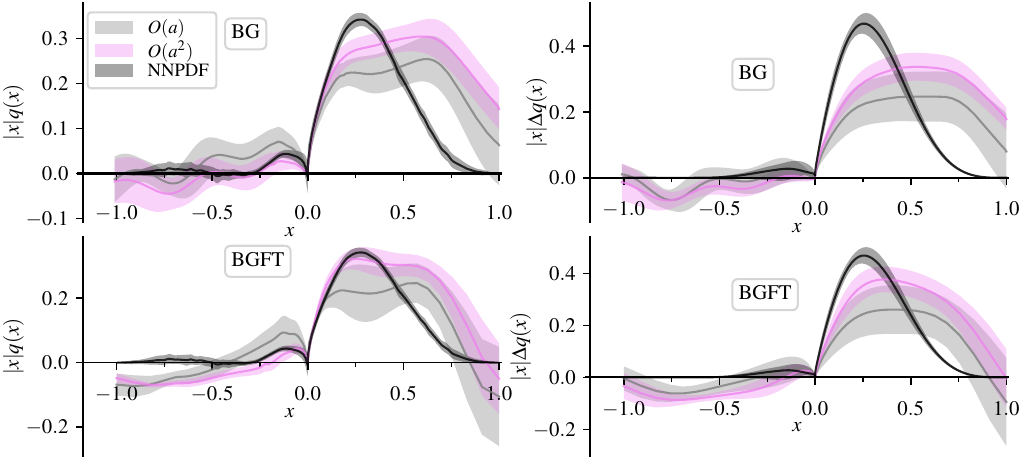}
  \caption{Unpolarized (left) and helicity (right) distributions in the continuum, using BG (top) and BGFT (bottom) methods. $O(a)$ and $O(a^2)$ extrapolations are shown in gray and pink, respectively. PDFs extracted through global fits from the releases NNPDF~\cite{Ball:2017nwa,Nocera:2014gqa} (dark gray) are included for qualitative comparison.}
  \label{fig:systematic_PDF}
\end{figure*}

In Fig.~\ref{fig:systematic_PDF}, we compare the distributions
obtained using $O(a)$ and $O(a^2)$ extrapolations with those obtained
from phenomenology by NNPDF~\cite{Ball:2017nwa, Nocera:2014gqa}. This
comparison is intended to be qualitative, since our calculation was
not done at the physical pion mass and does not include a study of
other sources of systematic uncertainty such as finite-volume effects
or the dependence on $p_z$.

In the antiquark region ($x<0$), the NNPDF result is slightly positive
for $x>-0.25$, particularly in the unpolarized case. Focusing on the
latter case, both of the extrapolations using both BG and BGFT methods
reproduce this feature, although the $O(a)$ extrapolation (which has a
larger uncertainty) prefers a wider and larger positive region. This
agreement with NNPDF is only present after the continuum extrapolation
and does not appear in the analyses of any of the individual
ensembles. For larger negative $x$, the NNPDF distributions are close
to zero. However, the BGFT result is below zero, particularly when
using an $O(a^2)$ extrapolation.

In the quark region ($x>0$), the distributions obtained from our data
tend to have smaller peaks at larger $x$ than phenomenology and fall
off more slowly at large $x$. All of the analyses are consistent with
zero at $x=1$, except for the $O(a^2)$-extrapolated BG data. For small
$x$, the lattice unpolarized distributions are consistent with
phenomenology, whereas the lattice helicity distributions have smaller
slopes. In the unpolarized case, the agreement holds for a wider range
of $x$ when using the BGFT approach, and this approach also produces
less disagreement in the helicity distribution.

\section{Conclusions}
\label{sec:conclusions}

In this work we performed a lattice QCD calculation of isovector
parton distributions via the quasi-PDF approach, using three twisted
mass ensembles with different lattice spacings. This enabled a study
of discretization effects, which can first appear at linear order in
the lattice spacing, and the approach to the continuum limit.

Although our data are unable to clearly distinguish $O(a)$ from
$O(a^2)$ contributions, we nevertheless observed significant
discretization effects, both in the position-space matrix elements and
in the final parton distributions. In the antiquark region, taking the
continuum limit produces a reasonable agreement with phenomenology.
Previous calculations, such as the one at the physical pion mass in
Ref.~\cite{Alexandrou:2019lfo}, have failed to reproduce the
phenomenological behaviour at small negative $x$; our work suggests
that discretization effects contribute significantly to this
discrepancy. At larger negative $x$, the agreement is better when
using the Backus-Gilbert method, although the uncertainty is also
larger. In the quark region, the continuum extrapolation also has a
significant effect, although large disagreements with phenomenology
remain. The latter is unsurprising, as we have not controlled other
sources of systematic uncertainty.

Going beyond the naïve truncated discrete Fourier transform, we have
compared two reconstruction techniques for obtaining quasi-PDFs from a
finite set of lattice data. We found that the
Bayes-Gauss-Fourier-Transform method produces a somewhat better
agreement with phenomenology in the quark region and worse agreement
in the antiquark region, although for the latter the Backus-Gilbert
method has a larger uncertainty. Given the uncontrolled systematic
effects, these observations should be treated with caution.

We have also compared two different approaches for nonperturbative
renormalization of the nonlocal operator $\cO_\Gamma(z)$. The
auxiliary-field approach tends to produce significantly larger
renormalized matrix elements than the whole-operator approach,
particularly at large $z$. In this work we chose to study PDFs using
the latter because its results are more precise, but it will be
important to continue studying different renormalization approaches to
understand their different systematics and whether they all produce
the same continuum limit.

While we have demonstrated the importance of discretization effects,
more work will be needed to understand the relative importance of
$O(a)$ and $O(a^2)$ effects in typical calculations. This could be
done by performing calculations using a wider range of lattice
spacings or by applying Symanzik improvement to remove $O(a)$
effects~\cite{Green:2020xco}.

\begin{acknowledgments}
  We thank Bartosz Kostrzewa for useful guidance concerning ensemble
  D45. K.C. and A.S. acknowledge support by the National Science Centre
  (Poland) grant SONATA BIS no.\ 2016/22/E/ST2/00013. 
  M.C. acknowledges financial support by the U.S. Department of Energy, 
  Office of Nuclear Physics, Early Career Award under Grant No.\ DE-SC0020405. 
  This project has received funding from the Marie Skłodowska-Curie
  European Joint Doctorate program STIMULATE of the European
  Commission under grant agreement No 765048; F.M. is funded under
  this program.
  This research used resources on the supercomputers
  JURECA~\cite{jureca} at Jülich Supercomputing Centre and Eagle at
  Poznań Supercomputing and Networking Center. For the RI-xMOM
  calculation, gauge fixing was performed using Fourier-accelerated
  conjugate gradient~\cite{Hudspith:2014oja}, implemented in
  GLU~\cite{GLU}. Calculations were performed using the Grid
  library~\cite{Boyle:2016lbp} and the DD-$\alpha$AMG
  solver~\cite{Frommer:2013fsa} with twisted mass
  support~\cite{Alexandrou:2016izb}. Data analysis and plotting were
  done using NumPy~\cite{numpy}, SciPy~\cite{scipy}, and
  Matplotlib~\cite{Hunter:2007}.
\end{acknowledgments}

\appendix

\bibliography{continuum_limit.bib}

\end{document}